\documentclass[a4paper,12pt]{article}
\pdfoutput=1

\usepackage{jheppub}

\usepackage{a4wide,epsfig,psfrag,amsmath,amssymb,scalefnt}
\usepackage[dvipsnames]{xcolor}
\usepackage{amsmath,comment,braket}
\usepackage{enumitem}
\usepackage{placeins}
\usepackage{breqn}
\usepackage{slashed}
\usepackage{comment}
\usepackage{subcaption}
\usepackage{hyperref}
\usepackage{url}
\usepackage{physics}

\usepackage{listings}
\allowdisplaybreaks

\begin{document}

\preprint{P3H-26-086, TTP26-039}

\title{Two-loop helicity amplitudes for $gg\to W^+W^-$ using high-energy and forward expansions}

\author{Mateusz Czaja,}
\author{Ben Rott,}
\author{Matthias Steinhauser,}
\author{Daniel Stremmer}

\emailAdd{
mateusz.czaja@kit.edu,
ben.rott@student.kit.edu,
matthias.steinhauser@kit.edu,
daniel.stremmer@kit.edu}

\affiliation{Institut f{\"u}r Theoretische Teilchenphysik, Karlsruhe Institute of Technology (KIT),  Wolfgang-Gaede Stra\ss{}e 1, 76131 Karlsruhe, Germany}

\abstract{

We calculate next-to-leading order QCD corrections for the
process $gg \to W^+W^-$ with two off-shell $W$ bosons, accounting for virtual top quarks in the loop
diagrams. Our approach relies on combining expansions valid in
different regions of phase space. Specifically, we perform a
high-energy expansion in which the Mandelstam variables are much
larger than the top-quark mass. Furthermore, we expand around the
forward-scattering limit, where the Mandelstam variable $t$ is
small. We also provide an
expansion for large top-quark mass, which serves as an important
cross-check.
We provide technical
details regarding the various expansions and demonstrate that precise
results are obtained for all helicity amplitudes across the entire
phase space. Our results exhibit an explicit dependence on all input
parameters.

}

\maketitle
\flushbottom

\newpage

\vspace*{2em}

\section{Introduction}

The production of vector-boson pairs is among the processes being
studied in detail at the Large Hadron Collider (LHC)~\cite{ATLAS:2025dhf,CMS:2024ild,ATLAS:2023dew,ATLAS:2026xuq,ATLAS:2022wnf,ATLAS:2021mbt}. They
are measured with a precision in the range of a few percent. This
applies in particular to $W^+ W^-$ production, which exhibits the
lowest relative uncertainty of 3.1\%~\cite{ATLAS:2025dhf}, followed by the $WZ$ channel at 3.4\%~\cite{CMS:2024ild}.
The continuum production of $W$ boson pairs is also an important background
for measurements of Higgs boson production with subsequent decay into
(off-shell) $W^+W^-$~\cite{Duhrssen:2005bz,Binoth:2005ua}.

Precise measurements require precise theoretical predictions. For the
total cross section of $W^+ W^-$ production, $q\bar{q}$ annihilation
is the dominant production channel; it is known up to next-to-next-to-leading order (NNLO)~\cite{Gehrmann:2014fva,He:2024iqg} and accounts for up to 95\%. The
remaining 5\% is attributed to the gluon fusion channel. The relative
weighting changes when considering cross sections in a restricted
phase space (fiducial cross sections). For instance, with cuts on the lepton opening angle, invariant mass and transverse momenta and a veto on jets, the gluon fusion channel can reach a share of 
approximately 25\%~\cite{Binoth:2005ua}.

It is known that the next-to-leading order (NLO) corrections for
gluon-induced processes are large, typically lying in the range between 50\% and
100\%.  This also applies to the massless contributions, which have been calculated in Refs.~\cite{Caola:2015rqy,Grazzini:2020stb} with the two-loop amplitudes from Refs.~\cite{Caola:2015ila,vonManteuffel:2015msa}. At LO, the gluon-induced corrections involving the top quark account for approximately 12\% and 15\% of the
massless contribution to the total and fiducial cross section respectively~\cite{Binoth:2006mf}. Therefore, it is important to know the NLO corrections for the top-quark contribution as well.

Besides fixed-order QCD correction also, the NNLO predictions have been
combined with parton showers
in the \texttt{MiNNLO}~\cite{Lombardi:2021rvg}
and the \texttt{GENEVA} frameworks~\cite{Gavardi:2023aco}.
More recently threshold logarithms have
been resummed including next-to-next-to-leading logarithms (NNLL) in Ref.~\cite{Banerjee:2025tbo}.
NLO electroweak corrections are
available in Refs.~\cite{Denner:2011id,Billoni:2013aba,Bierweiler:2012kw,Biedermann:2016guo}.

In this work, we consider the top-quark-induced contributions to $gg\to
W^+W^-$. The corresponding helicity amplitudes have already been
calculated in Ref.~\cite{Bronnum-Hansen:2020mzk}, albeit using
exclusively numerical methods. In particular, this approach requires
fixing the input values at an early stage of the calculation. Our
calculation yields helicity amplitudes that depend explicitly on all
input parameters. They can be easily implemented into numerical code
(e.g., the \texttt{C++} library \texttt{ggxy}~\cite{Davies:2025qjr}), offering a
high degree of flexibility
and an interface to \texttt{POWHEG}~\cite{Nason:2004rx,Frixione:2007vw,Alioli:2010xd}. Furthermore,
we consider off-shell $W$ bosons, i.e., we introduce two different
virtualities for the $W$ bosons in the final state, which allows the straightforward inclusion of $W$-boson decays.

\begin{figure}[b]
\centering
\includegraphics[width=1.0\textwidth]{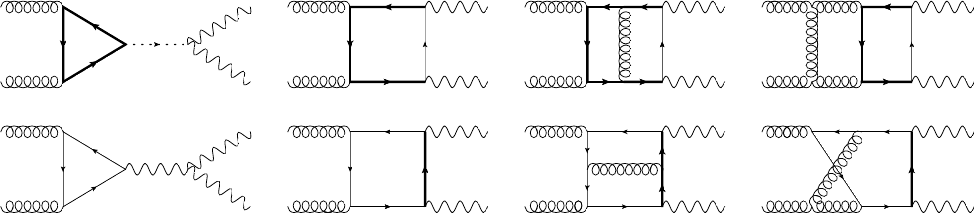}
\caption{\label{fig::FDs}
  Sample Feynman diagrams contributing to $gg\to W^+W^-$.
  Thick and thin solid lines represent massive top quarks and massless bottom quarks, respectively. Curly, wavy and dashed lines
  stand for gluons, $Z$ bosons or photons and Higgs bosons, respectively.
  }
\end{figure}

In Fig.~\ref{fig::FDs}, we show representative one- and two-loop
Feynman diagrams contributing to $gg\to W^+W^-$. Top and bottom quarks
appear in the loops of the box diagrams; the bottom quarks are assumed to be
massless. Additionally, there are triangle diagrams mediated by a Higgs, a $Z$ boson or a photon. The contribution from the photon and the
contribution from the vector coupling of the $Z$ boson vanish due to
Furry's theorem~\cite{Furry:1937zz}, while the Higgs boson
contribution can be derived from the corresponding calculation for
$gg\to ZZ$~\cite{Davies:2025out}. 

Regarding the triangle 
contribution involving the axial-vector coupling of the $Z$ boson, we find that it vanishes
for on-shell $W$ bosons, whereas for off-shell $W$ bosons it yields a
non-zero result that depends on the gauge parameter, see also Ref.~\cite{Binoth:2006mf}. It should be
noted that this gauge-parameter dependence cancels out when
considering physical processes -- such as those where the $W$ bosons
decay leptonically.
For this reason, we will not consider this contribution in the present
work, but rather postpone it until the process $gg\to W^+W^-$ is
implemented in \texttt{ggxy}.

The remainder of the paper is structured as follows: In the next
section, we explain the technical details of our calculation, paying
particular attention to the analytical determination of the master
integrals in the high-energy and forward-scattering limits. Due to the
simultaneous appearance of top and bottom quarks
within the loop diagrams, most of these
integrals have not previously been considered in the literature
in the context of high-energy or forward expansions. 
We demonstrate our approach to combine the $t$ and high-energy expansions
for top-level two-loop scalar integrals in Section~\ref{sec::masters}.
The
results for the helicity amplitudes are discussed in
Section~\ref{sec::results}; we combine the various components and
demonstrate that precise results are obtained across the entire phase
space. Finally, a summary is provided in Section~\ref{sec::concl}.
In Appendix~\ref{app::sym} we present all
symmetry relations between the helicity amplitudes,
in Appendix~\ref{app::largemt} we discuss the large
top quark mass limit of the helicity amplitudes, and in Appendix~\ref{app::topos} we define the two example topologies we used to investigate the convergence of the $t$ and the high-energy expansions.

\section{Technical details}

This section is dedicated to the technical aspects of our calculation. In particular, we define the helicity amplitudes
used to calculate the virtual NLO corrections
and describe the calculation of the helicity amplitudes in the
forward and high-energy limits. We restrict the discussion
primarily to the differences compared to the calculations
for $gg\to HH$~\cite{Davies:2019dfy,Davies:2023vmj}, $gg\to ZH$~\cite{Davies:2025out}, and $gg\to ZZ$~\cite{Davies:2025out}, which
were performed according to the same principle.
As we will see, due to the presence of the (massless) bottom quark, the different decompositions of the masses
inside the loop diagrams lead to different integral families and
consequently to new master integrals.

We perform the standard renormalization of the ultra-violet
divergences and the subtraction of the infra-red poles
and refer to Section 2.4 of Ref.~\cite{Davies:2025out}
for more details.

\subsection{Kinematics}

We consider the process $g(q_1) g(q_2) \to W^+(q_3) W^-(q_4)$
where all momenta are incoming with $q_1^2=q_2^2=0$.
We keep the invariants $q_3^2$ and $q_4^2$ arbitrary to
allow for off-shell $W$ boson production.
The Mandelstam variables are defined as 
\begin{eqnarray}
  s = (q_1 + q_2)^2\,, \nonumber\\
  t = (q_1 + q_3)^2\,, \nonumber\\
  u = (q_2 + q_3)^2\,,
\end{eqnarray}
which leads to the relation $s+t+u=q_3^2+q_4^2$.

\subsection{Amplitude construction}

The helicity amplitudes for $gg\to W^+W^-$ can be written as
\begin{align}
\mathcal{A}^{\omega}_{\lambda_1\lambda_2\lambda_3\lambda_4}(q_1,q_2,q_3,q_4) = \mathcal{A}_{\mu\nu\rho\sigma}^{\omega}(q_1,q_2,q_3,q_4)\epsilon_{\lambda_1}^\mu(q_1)\epsilon_{\lambda_2}^\nu(q_2)\epsilon_{\lambda_3}^{*\rho}(q_3)\epsilon_{\lambda_4}^{*\sigma}(q_4)\,,
\end{align}
where the polarization indices $\lambda_i$ take the values $\{+,-\}$ for gluons and $\{+,-,0\}$ for $W$ bosons. The superscript $\omega={v,a}$ indicates the vector and axial-vector contributions. The structure of the vector contribution is identical to the one of $gg\to Z^\star Z^\star$ and we use the projectors from Ref.~\cite{Davies:2025out}. For the axial-vector contribution we follow the approach of Ref.~\cite{Gehrmann:2023zpz} to construct a basis of tensor structures. In particular, we define the vector
\begin{align}
v_A^\mu=\epsilon^{\mu\nu\rho\sigma}q_{1,\nu}q_{2,\rho}q_{3,\sigma},
\end{align}
and the tensor
\begin{align}
\mathcal{T}_O^{\mu\nu\rho}=g^{\mu\nu}v_A^\rho+g^{\mu\rho}v_A^\nu+g^{\nu\rho}v_A^\mu,
\end{align}
so that all parity-odd rank $4$ tensors can be constructed from the combination of the external momenta $\{q_1^\mu,q_2^\mu,q_3^\mu\}$ and either $v_A^\mu$ or $\mathcal{T}_O^{\mu\nu\rho}$. 
In this way we find $120$ parity-odd tensor structures that can be reduced to the following $18$ 
\begin{alignat}{4}
  &T_{1}^{\mu\nu\rho\sigma}=q_1^\rho q_3^\mu q_3^\nu v_A^\sigma,\quad
  &&T_{2}^{\mu\nu\rho\sigma}=q_2^\rho q_3^\mu q_3^\nu v_A^\sigma,\quad
  &&T_{3}^{\mu\nu\rho\rho}=q_1^\sigma q_3^\mu q_3^\nu v_A^\nu,\quad
  &&T_{4}^{\mu\nu\rho\rho}=q_2^\sigma q_3^\mu q_3^\nu v_A^\nu,\quad
  \nonumber\\
  &T_{5}^{\mu\nu\rho\nu}=q_1^\rho q_1^\sigma q_3^\mu v_A^\nu,\quad
  &&T_{6}^{\mu\nu\rho\sigma}=q_1^\rho q_2^\sigma q_3^\mu v_A^\nu,\quad
  &&T_{7}^{\mu\nu\rho\sigma}=q_1^\sigma q_2^\rho q_3^\mu v_A^\nu,\quad
  &&T_{8}^{\mu\nu\rho\sigma}=q_2^\rho q_2^\sigma q_3^\mu v_A^\nu,\quad
  \nonumber\\
  &T_{9}^{\mu\nu\rho\sigma}=\mathcal{T}_O^{\nu\rho\sigma}q_3^\mu,\quad
  &&T_{10}^{\mu\nu\rho\sigma}=q_1^\rho q_1^\sigma q_3^\nu v_A^\mu,\quad
  &&T_{11}^{\mu\nu\rho\sigma}=q_1^\rho q_2^\sigma q_3^\nu v_A^\mu,\quad
  &&T_{12}^{\mu\nu\rho\sigma}=q_1^\sigma q_2^\rho q_3^\nu v_A^\mu,\quad
  \nonumber\\
  &T_{13}^{\mu\nu\rho\sigma}=q_2^\rho q_2^\sigma q_3^\nu v_A^\mu,\quad
  &&T_{14}^{\mu\nu\rho\sigma}=\mathcal{T}_O^{\mu\rho\sigma}q_3^\nu,\quad 
  &&T_{15}^{\mu\nu\rho\sigma}=\mathcal{T}_O^{\mu\nu\sigma}q_1^\rho,\quad
  &&T_{16}^{\mu\nu\rho\sigma}=\mathcal{T}_O^{\mu\nu\sigma}q_2^\rho,\quad
  \nonumber\\
  &T_{17}^{\mu\nu\rho\sigma}=\mathcal{T}_O^{\mu\nu\rho}q_1^\sigma,\quad
  &&T_{18}^{\mu\nu\rho\sigma}=\mathcal{T}_O^{\mu\nu\rho}q_2^\sigma,
  &&
  &&
  \label{eq::Ti}
\end{alignat}
by using the transversality of the external particles ($\varepsilon(q_i)\cdot q_i=0$) and the gauge conditions $\varepsilon(q_1)\cdot q_2=\varepsilon(q_2)\cdot q_1=0$ for the two gluons. Based on this independent set of tensor structures we construct the projectors to the axial-vector helicity amplitudes following the same approach as in the vector case described in Ref.~\cite{Davies:2025out}.

Due to different symmetry relations 
(see Appendix~\ref{app::sym})
it is sufficient to consider only eight 
among the helicity amplitudes for the vector and axial-vector part, respectively, which we choose as
\begin{eqnarray}
    \mathcal{A}^{\omega}_{++++},\, \mathcal{A}^{\omega}_{+++-},\, \mathcal{A}^{\omega}_{++-0},\, \mathcal{A}^{\omega}_{++00},\, 
    \mathcal{A}^{\omega}_{+-++},\, \mathcal{A}^{\omega}_{+-+-},\, \mathcal{A}^{\omega}_{+--0},\, \mathcal{A}^{\omega}_{+-00}\,.
     \label{eq::helamp}
\end{eqnarray}
The perturbative expansion of the helicity amplitudes is written as 
\begin{equation}
\mathcal{A}^{\omega}_{\lambda_1\lambda_2\lambda_3\lambda_4} = \frac{\alpha_s}{2\pi}\mathcal{A}^{(0),\omega}_{\lambda_1\lambda_2\lambda_3\lambda_4}+\left(\frac{\alpha_s}{2\pi}\right)^2\mathcal{A}^{(1),\omega}_{\lambda_1\lambda_2\lambda_3\lambda_4}.
\end{equation}

\subsection{\label{sub::he}High-energy expansion}

In the following we discuss the expansion of the helicity amplitudes in the high-energy region, where we consider the top-quark mass to be larger than the masses of the two external $W$ bosons but smaller than the Mandelstam variables, i.e., we have
\begin{eqnarray}
  q_3^2,q_4^2\ll m_t^2\ll s,t,u\,. 
\end{eqnarray}
A similar approach has been used in Refs.~\cite{Davies:2019dfy,Davies:2023vmj,Davies:2025out}
for the processes
$gg\to HH$, $gg\to ZZ$, $gg\to ZH$.

The first step of this calculation is the expansion of the amplitudes in the limit $q_3^2,q_4^2\to 0$, which corresponds to a Taylor expansion. The expansion of the scalar integrals is carried out with {\tt LiteRed}~\cite{Lee:2013mka}. The results are expressed in terms of scalar integrals with $q_3^2=q_4^2=0$. 
Before inserting the expansions into the amplitude
it turns out to be convenient to apply several simplifications.
This involves an integration-by-parts (IBP) reduction,
which we perform with {\tt Kira3}~\cite{Lange:2025fba} and {\tt FireFly}~\cite{Klappert:2019emp,Klappert:2020aqs}, and 
a partial fraction decomposition in $\epsilon$ using 
{\tt FORM}~\cite{Davies:2026cci} procedures generated with {\tt MultivariateApart}~\cite{Heller:2021qkz}. In this process
it is important to work with a basis where the dependence on $\epsilon$ and the kinematic variables factorizes. This is
achieved with the help of 
{\tt ImproveMasters.m}~\cite{Smirnov:2020quc}.
We manage to express our amplitude in terms of
$254$ master integrals of which only $18$ are already known
from previous calculations~\cite{Davies:2019dfy,Davies:2023vmj,Davies:2025out}.
In particular, all master integrals in the top-level sector with
seven propagators are new.

A challenging part of our calculation is the computation
of the master integrals in the high-energy limit
where $s,t\gg m_t^2$. We profit a lot from previous
calculations~\cite{Davies:2019dfy,Davies:2023vmj,Davies:2025out} and proceed along the same
lines. As previously mentioned, however, the majority of our master integrals are new, which compelled us to further refine the methods used to date.

Differential equations form a crucial basis for calculating the master integrals.
From the three dimensionful quantities of our problem,
we form two ratios: $m_t^2/s$ and $t/s$. Using IBP reduction tables,
two systems of differential equations can thus be set up.
We use the system in $m_t^2/s$ to perform the expansion
for small top-quark mass. The differential equation in
$t/s$ is useful for determining the corresponding boundary conditions.
We want to stress that it is not necessary
to bring the differential equations into
a particular from. For example, it is not
necessary that the $\epsilon$ dependence 
factorizes from the differential equation matrix.

For the computation in the limit $m_t\to 0$ we insert
for each master integral
the following ansatz  
into the differential equations with respect to $m_t^2/s$
\begin{align}\label{eq:ansatz-he}
    M_i(s,t,m_t,\epsilon)=\sum_{a=-4}^{a_{i,{\rm max}}}\sum_{b=-6}^{b_{i,{\rm max}}}\sum_{c=0}^{4+a}c_{abc}^{(i)}(s,t)\epsilon^a \left(\frac{m_t}{\sqrt{s}}\right)^b 
    \log^c\left(\frac{m_t^2}{s}\right)\,.
\end{align}
The value of the upper limit $a_{i,{\rm max}}$ depends on the spurious poles in front of the respective master integral and $b_{i,{\rm max}}$ on the desired expansion depth, which in our case is $120$. This value is sufficiently large such that after accounting for spurious $m_t$ poles in the prefactors of the master integrals we obtain $100$
expansion terms for each helicity amplitude.

After expanding the coefficients of the differential equations in the limit $m_t\to 0$ and $\epsilon\to 0$ we obtain a system of linear equations for the functions $c_{abc}^{(i)}(s,t)$ which we solve with {\tt Kira}
using the option to provide a user-defined system.
This allows us to write all master integrals in terms of 680 boundary functions $c_{abc}^{(i)}(s,t)$. Next we insert this result for the master integrals into the differential equations with respect to $t$ to derive a system of differential equations with respect to $t/s$ of the boundary conditions $c_{abc}^{(i)}(s,t)$. In this way it is sufficient to compute the functions $c_{abc}^{(i)}(s,t)$ in the limit $t\to 0$ and the full $t$ dependence can be obtained from solving the associated differential equations with respect to $t$.

For the calculation of the boundary conditions ($c^{(i)}_{abc}(s,t)$) we apply the method of regions~\cite{Beneke:1997zp} to expand the master integrals in the limit $m_t\to 0$. For this computation we make extensive use of the \texttt{Mathematica} package \texttt{AsyInt} \cite{Zhang:2024fcu}, where several steps of the calculation, such as the asymptotic expansion using so-called template integrals and some of the further handling of Mellin-Barnes integrals, have been automated. In practice we modify some of the low-level functions in \texttt{AsyInt} to enhance the parallel computing power in \texttt{Mathematica}.

The asymptotic regions are obtained with the {\tt Mathematica} package {\tt asy2.1.m} \cite{Jantzen:2012mw}, where we consider the following scalings
\begin{align}
    m_t^2\sim \rho,{\qquad} \{S,T,U\}\sim 1.
\end{align}
We note that {\tt asy2.1.m} requires as input positive definite Symanzik polynomials, where we have to use the Euclidean Mandelstam invariants $\{S,T,U\}=\{-s,-t,-u\}$ with $S,T,U>0$. However, we are allowed to use $S+T+U=0$
to simplify the Symanzik polynomials. Note that
we have to keep the exact dependence on $U$.
The output is a list of scalings of the $\alpha_i$ parameters for each region.
In our application the $\alpha_i$ can scale as $1$, $\rho$ or $\sqrt{\rho}$ (for non-planar integrals only). After expanding the integrand for $\rho\to 0$ and performing the analytic continuation $\epsilon\to 0$ with \texttt{MB.m}~\cite{Czakon:2005rk}, the master integrals are given by a power-log series as shown in Eq.~\eqref{eq:ansatz-he}, where the function $c_{abc}^{(i)}(s,t)$ can be written in terms of Mellin-Barnes integrals. 

We encounter various types of Mellin-Barnes integrals, which we discuss below. For the simpler cases, we evaluate the Mellin-Barnes integrals directly and obtain the full $t$-dependence. The more complex Mellin-Barnes integrals are expanded in the limit $t \to 0$; this serves as a boundary condition for the $t$-differential equations.

The various types of Mellin-Barnes
integrals are:
\begin{itemize}
\item
One-dimensional Mellin-Barnes integrals that are evaluated numerically to high precision with \texttt{MB.m}, and the analytical result is reconstructed using the PSLQ algorithm~\cite{PSLQ}. 
\item 
We also encounter one-dimensional Mellin-Barnes integrals with a single scale of the form
\begin{align}
    \int^{+i\infty}_{-i\infty}\frac{\dd z}{2\pi i} x^z f(z),
    \label{eq::MB1}
\end{align}
where $f(z)$ is a rational expression in terms of Gamma and PolyGamma functions and $x\in\{T/S,U/S\}$. We compute most of these integrals analytically by summing over residues following Cauchy's theorem. In this context we find it useful to use the tools \texttt{HarmonicSums.m}~\cite{Blumlein:1998if,Vermaseren:1998uu,Blumlein:2009ta,Ablinger:2009ovq,Ablinger:2011te,Ablinger:2012ufz,Ablinger:2013eba,Ablinger:2013cf,Ablinger:2014bra,Ablinger:2014rba,Ablinger:2015gdg,Ablinger:2018cja} \texttt{Sigma.m}~\cite{sigmaI,sigmaII} and \texttt{EvaluateMultiSums.m}~\cite{RISC4711,RISC4826}.
In some other cases 
we expand the Mellin Barnes representation of
Eq.~(\ref{eq::MB1}) in the limit $T\to 0$ with \texttt{MBasymptotics}~\cite{MBasymtotics}. 

\item 
Two-dimensional Mellin-Barnes integrals with one or two scales are present in six- and seven-line master integrals. We expand them in the limit $T\to 0$. This expansion leads to several one-dimensional Mellin-Barnes integrals which are calculated as discussed above. 

\item
In practice we also find two-dimensional Mellin-Barnes integrals with a single scale $U/S$, which we cannot expand. The results for the master integrals that contain such integrals are obtained from crossing relations ($t\leftrightarrow u$) at the level of the master integrals. 
This means, whenever we encounter such a two-dimensional Mellin-Barnes
integral it is possible to choose a master integral where $t$ and $u$ are interchanged and which thus depends on $T/S$.

\item 
We also find three-dimensional Mellin-Barnes integral with two scales, which we also expand in the limit $T\to 0$. This leads to one- and two-dimensional Mellin-Barnes integrals. The two-dimensional Mellin-Barnes integrals always contain the scale $y=U/S$ so that no further expansion is possible. In practice we solve such Mellin-Barnes integrals with the \textit{expand and fit method}, see Refs.~\cite{Mishima:2018olh,Zhang:2024fcu}.
In this approach one first computes
deep expansions of the Mellin-Barnes integrals in the limit $y\to 0$. In the most difficult cases we need expansions up to
$\sim y^{300}$. Afterwards, one performs a fit to an ansatz that is constructed from Harmonic PolyLogarithms (HPLs) up to weight~5 with the letters $\{1/y,1/(1+y)\}$, compromising $32$ basis elements. For our calculation, each of them is multiplied by a linear combination of up to nine monomials from $\{1,1/y,1/y^2,1/y^3,1/(1+y),1/(1+y)^2,1/(1+y)^3,1/(1+y)^4,1/(1+y)^5\}$.

We observe that the three-dimensional Mellin-Barnes integrals appear
only in the $\mathcal{O}(\epsilon)$ terms of the non-planar top-level
master integrals. In principle, the calculation of such
terms could be avoided by choosing an alternative basis for the master
integrals. However, since we were unable to find such a basis, we
regard the cancellation of weight-5 objects in the final
results for the helicity amplitudes as a welcome consistency check.

\end{itemize}

For some of the master integrals we also use
an alternative method to compute the high-energy
expansion.
Here we do not perform an asymptotic expansion for
small top-quark mass using the method of regions. Consequently, we do not use \texttt{asy2.1} and do not have to introduce additional regulators which are required in the method of regions (see the template integral approach introduced in Ref.~\cite{Mishima:2018olh}). Instead, we derive a Mellin-Barnes
representation keeping the full dependence on $m_t$
using the \texttt{Mathematica} package \texttt{MBcreate}~\cite{Belitsky:2022gba}. Subsequently, the Mellin-Barnes integrals can be expanded in small $m_t$ with \texttt{MBasymptotics}. We find that for our calculation, considering integrals with six or fewer lines, this approach is more efficient. For the most complicated seven-line integrals, if the full dependence on $m_t$ is kept, the Mellin transforms generated by \texttt{MBcreate} lead to high-dimensional Mellin-Barnes integrals. In certain cases, these can be reduced with the application of Barnes lemmas. When this proved not possible, we relied on the standard method of regions approach discussed above.

\subsection{Forward expansion} 

For the expansion in the forward limit we follow the strategy used and
outlined in Refs.~\cite{Davies:2023vmj,Davies:2025out}. In fact we follow two approaches
which we briefly describe in the following.

In the first approach we perform an expansion of the integrand in the
variable $\delta q = q_1 + q_3$.  This corresponds to a simultaneous
Taylor expansion in $t$, $q_3^2$ and $q_4^2$. As a consequence, the resulting
integrals only depend on $s$ and $m_t^2$. After partial fraction decomposition, we arrive at new integral families for which we perform a reduction to $94$ master integrals
and derive a differential equation in $s/m_t^2$.

The master integrals we encounter in our calculation are different
from those from the previous calculation with neutral bosons in the
final state~\cite{Davies:2023vmj,Davies:2025out}. Due to the $tbW$ coupling the massive
line is not through-going which leads to simpler integrals. In fact,
whereas in the previous cases there was an elliptic
sector~\cite{Bonciani:2018uvv,Davies:2023vmj,Davies:2025out} all master integrals in the
current calculation can be computed analytically in terms of iterated
integrals. The results are obtained with the help of the differential
equations and boundary conditions computed in the large-$m_t$ limit. The differential equations contain two square roots, $\sqrt{x}\sqrt{4-x}$ and $\sqrt{x}\sqrt{4+x}$, where the latter appears only in two master integrals and cancels completely in the final amplitude.

We compute all master integrals also with the help of \textit{expand
and match}~\cite{Fael:2021kyg,Fael:2021xdp,Fael:2022miw}. This is an algorithmic procedure which
leads to deep (about 50 terms) generalized expansions around
chosen expansion points with numerical coefficients. Boundary conditions are either fixed with the large-$m_t$ limit or with numerical results obtained with \texttt{AMFlow}~\cite{Liu:2022chg}.
We manage to determine the master integrals such that we 
obtain a precision of at least 10 digits at the level of the
helicity amplitudes for all values of $s/m_t^2$.

We compare our analytical and numerical results for the master
integrals in detail. For the numerical evaluation of the helicity
amplitudes and particularly for their implementation in
\texttt{ggxy} it proves more efficient to use the numerical version
based on the \textit{expand and match} method.

Among the technical difficulties in this approach are the tensor
integrals which appear after the expansion in $\delta q$. For the leading
term we already have tensor integrals of rank~$2-6$ depending on the helicity amplitude  which increases up to
rank~$16$ for the fourth/fifth-order\footnote{Three helicity amplitudes are expanded to fourth and five to fifth order, see below.} term in $t$ and $q_{3,4}^2$. Furthermore we
observe a rapid increase in the number of terms for each further
expansion term.
To perform the tensor reduction in \texttt{FORM}~\cite{Davies:2026cci}
we use the package \texttt{OPITeR}~\cite{Goode:2024cfy} to generate templates for the specific ranks of the tensors.

Using this approach we manage to compute expansion terms 
for all helicity amplitudes up
to fourth order, i.e., all expansion terms up
to $t^n (q_3^2)^m (q_4^2)^k$ with $n+m+k=4$ and $n,m,k\ge 0$.
For the helicity amplitudes $\mathcal{A}_{++++},\mathcal{A}_{++-0},\mathcal{A}_{++00},\mathcal{A}_{+-+-},\mathcal{A}_{+--0}$ we manage to obtain results with  $n+m+k=5$. In the case of the other three helicity amplitudes $\mathcal{A}_{+++-},\mathcal{A}_{+-++},\mathcal{A}_{+-00}$ the leading expansion term vanishes and it is possible to factor out an overall prefactor of $p_T^2$, where $p_T$ is the transverse momentum of one of the $W$ bosons, such that the expansion can be written as $p_T^2\,t^n (q_3^2)^m (q_4^2)^k$ with $n,m,k\ge 0$. Therefore, the expansion up to $n+m+k=4$ of these helicity amplitudes is of similar complexity and the same expansion depth when considering the scaling $p_T^2\sim t,q_{3,4}^2$.

In the second approach we use the same starting point as for the high-energy expansion, see Section~\ref{sub::he}, and perform a Taylor expansion in the final-state
virtualities up to sixth-order terms in $q_3$ and $q_4$, so that we end up with the same amplitude expressed in terms of the $254$ master integrals. Next, we expand the master integrals in the limit $t\to 0$, which is realized via a Taylor
expansion in $t$ that is inserted into the $t$ differential equation. The boundary conditions are fixed by an expansion of the master integrals in $\delta q=q_1+q_3$. Since we only expand individual
integrals we can construct a deeper expansion. We manage to obtain terms up to $t^{10}$ at the level of helicity amplitudes, i.e. for all helicity amplitudes we have an expansion up to $t^n (q_3^2)^m (q_4^2)^k$ with $n\le10$ and $m+k\le 3$.

For our final result we combine the results from the two
approaches. The deep expansion in $q_{3,4}^2$ obtained from the first approach is important for lower values of $s$ and $t$.

\section{\label{sec::masters}Results for two-loop scalar integrals}

\begin{figure}[t]
\centering
\includegraphics[width=0.8\textwidth]{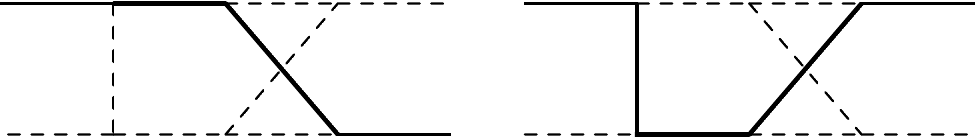}
\caption{\label{fig::MIs}
Top-level scalar integrals with seven propagators.
Left: \texttt{ggwwfull2l8(1,1,1,1\allowbreak{},1,1,1,0,-1)};
right: \texttt{ggwwfull2l11(1,1,1,1,1,1,1,0,0)}. 
Topology definitions of the two scalar integrals can be found in Appendix \ref{app::topos}.
Thick solid  and dashed internal lines represent massive
(mass $m_t$) and massless scalar propagators. 
Solid and dashed external lines represent virtualities
$q_3^2=q_4^2=m_W^2$ and $0$, respectively.
}
\end{figure}

\begin{figure}[t]
    \begin{center}
    \begin{tabular}{c}
    \includegraphics[width=.95\textwidth]{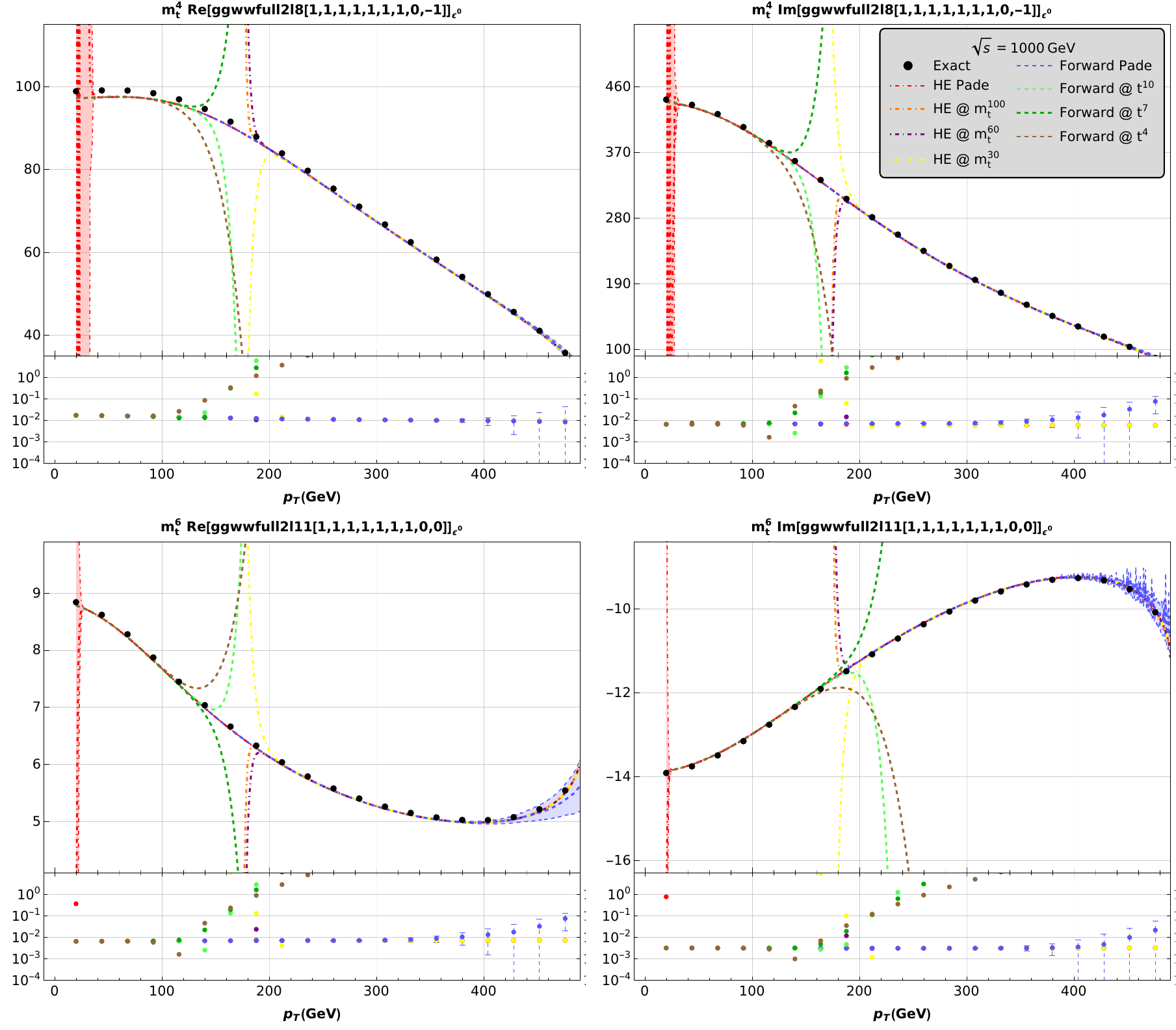} 
    \end{tabular}
    \end{center}
  \caption{\label{fig::ampl1lv_sqrts1000_HE}
  Real and imaginary parts of the integrals shown in Fig.~\ref{fig::MIs}. The integrals were rescaled by powers of $m_t$ to make them dimensionless. Results for the high-energy and the forward expansions are
  shown for different expansion depths in $m_t$ an $t$ respectively. The Pad\'e-improved results are compared to the exact numerical results (black dots).}
\end{figure}

\begin{figure}[t]
    \begin{center}
    \begin{tabular}{c}
    \includegraphics[width=.95\textwidth]{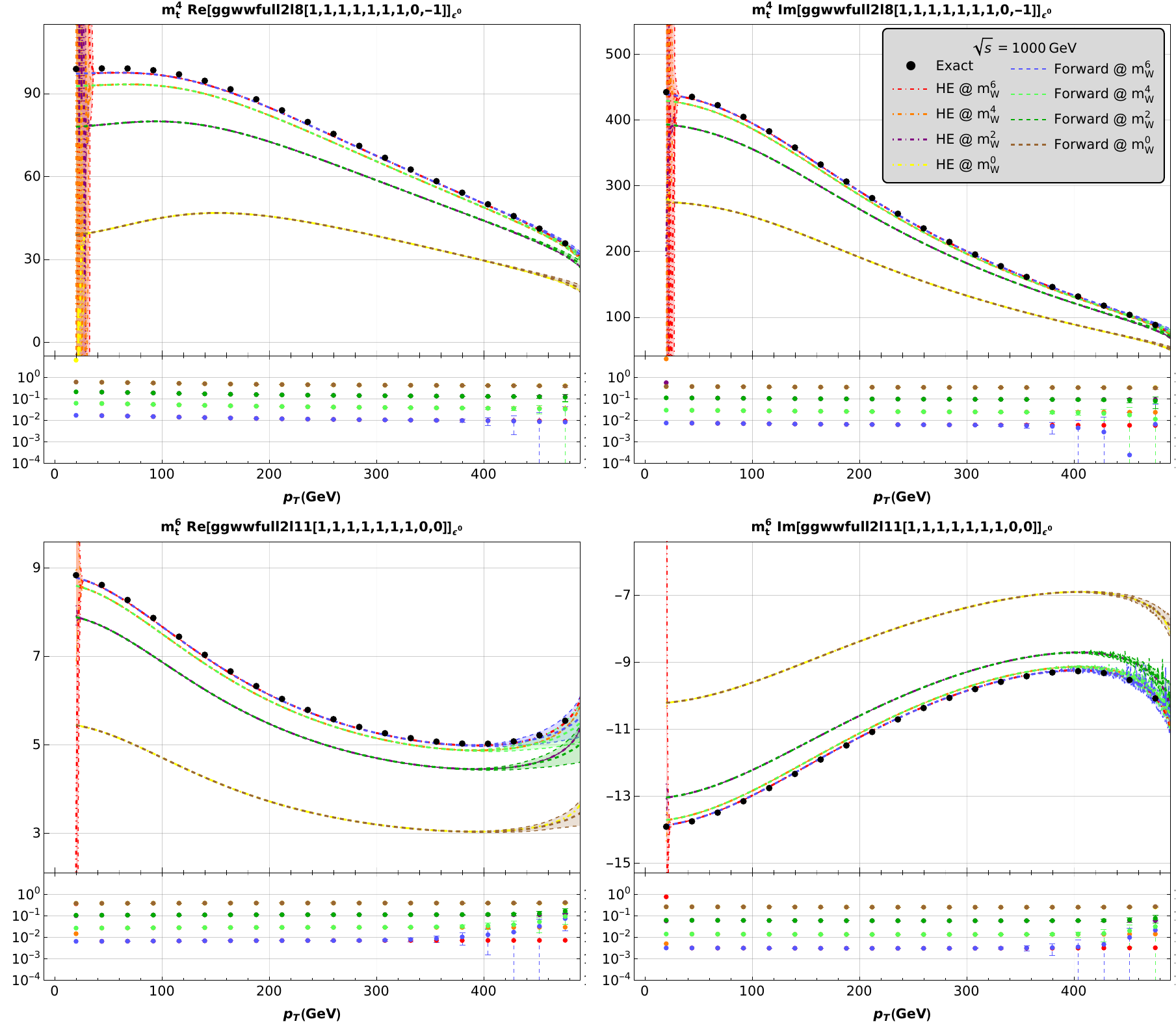} 
    \end{tabular}
    \end{center}
  \caption{\label{fig::ampl1lv_sqrts1000_q34}
    Real and imaginary parts of the master integrals shown in Fig.~\ref{fig::MIs}. The integrals were rescaled by powers of $m_t$ to make them dimensionless. Results for the forward and
    high-energy expansions  are
  shown for different expansion depths in the external virtualities. 
  Exact numerical results are shown as black dots.
  }
\end{figure}

In this Section we discuss the quality of our expansions 
for two sample top-level scalar integrals, which are
given in graphical form in Fig.~\ref{fig::MIs}.
As an example we choose $\sqrt{s}=1000$~GeV and discuss real
and imaginary parts as a function of the transverse momentum $p_T$.
As input we use $m_t=172.52$~GeV and
$q_3^2=q_4^2=m_W^2=(80.36)^2$~GeV$^2$.
We compare to exact results obtained with the help of
\texttt{AMFlow}~\cite{Liu:2022chg}.

Our results for the $t$-expansion and the high-energy expansion
include terms up to order $t^{10}$ and $m_t^{100}$, respectively. In
both cases, we improve the convergence by calculating all possible Pad\'e
approximants for a given range of expansion depths of the small parameter ($t$ or
$m_t$); that is, we rewrite the polynomial expansion in $t$ or $m_t$
as a rational function of the respective variable. Subsequently, a
weighted average is computed based on the following criteria: the
distance to the pole in the denominator closest to the physical point,
the number of expansion terms used to construct the Padé approximant,
and the deviation from the diagonal (i.e., from the equality of the
degrees of the numerator and denominator polynomials). This yields a
central value and an uncertainty -- both of which are shown in the plots
below. Further details regarding the algorithm can be found in Ref.~\cite{Davies:2020lpf}.

In Fig.~\ref{fig::ampl1lv_sqrts1000_HE} we present the convergence of the high-energy and the forward expansions, where in both cases we include expansion terms in the external masses up to $q_{3,4}^6=m_W^6$. 
For the high-energy region, we show
results expanded up to order $m_t^{30}$, $m_t^{60}$ and $m_t^{100}$,
respectively, and observe that they diverge around $p_T=200$~GeV. If
we apply the Pad\'e approximation in the variable $m_t$, which provides for each phase-space
point both a central value and an uncertainty estimate, we observe
that the agreement with the exact numerical result
is extended down to about $p_T=40$~GeV.
For smaller values of $p_T$, the uncertainty associated with the Pad\'e method increases significantly, as can be seen from the red bands on the left side of the panels.
Correspondingly, the Taylor expansion in $t$ is shown at the $t^4$, $t^7$, and $t^{10}$ orders. We note that the convergence of the forward expansion for these scalar integrals is worse compared to the ones of $gg\to HH$ \cite{Davies:2023vmj}, where typically a convergent behaviour is found up to $p_T\approx 200$~GeV. In $gg\to W^+W^-$ the convergence of the forward expansion is limited due to the scalar integrals involving $t$-channel cuts through a single massive top-quark line, while in $gg\to HH$ we always encounter two massive top-quark lines through $t$-channel cuts. On the other hand, our results when Pad\'e accelerated in the expansion parameter $t$, provide a broad overlap region between the two expansions, extending to $p_T=400$ GeV or higher for both integrals we investigated. 
The uncertainty of the Pad\'e procedure is represented as  bands on the right side of the panels. Overall,
we observe an agreement with the exact results at the level of 1\% or better for the whole parameter space.

In Fig.~\ref{fig::ampl1lv_sqrts1000_q34} we present the convergence in the external virtualities of the two integrals. We use the Pad\'e accelerated results for both the high-energy and the forward expansion. The series in $q_3^2$ and $q_4^2$ 
was truncated at the zeroth, first, second and third
orders. For the best approximation, we again find an agreement between the two expansions and the exact results at the level of 1\%. In addition, the agreement between the two expansion methods at the same order in the external-mass expansion provides a useful cross-check.


\section{\label{sec::results}Results for helicity amplitudes}

We dedicate this Section to the discussion
of the helicity amplitudes. In particular,
we show that the combination of the 
forward and high-energy expansion
covers the whole phase space. 
We demonstrate our approach to combine the $t$- and high-energy
expansions for the one-loop amplitudes in
Section~\ref{sub::1loop} where we can compare the results
with exact expressions. We discuss the two-loop
helicity amplitudes in Section~\ref{sub::2loop}.
We use the same input values as those specified at the beginning of Section~\ref{sec::masters}.

\subsection{\label{sub::1loop}One-loop amplitude}

\begin{figure}[t]
    \begin{center}
    \begin{tabular}{cc}
    \includegraphics[width=.45\textwidth]{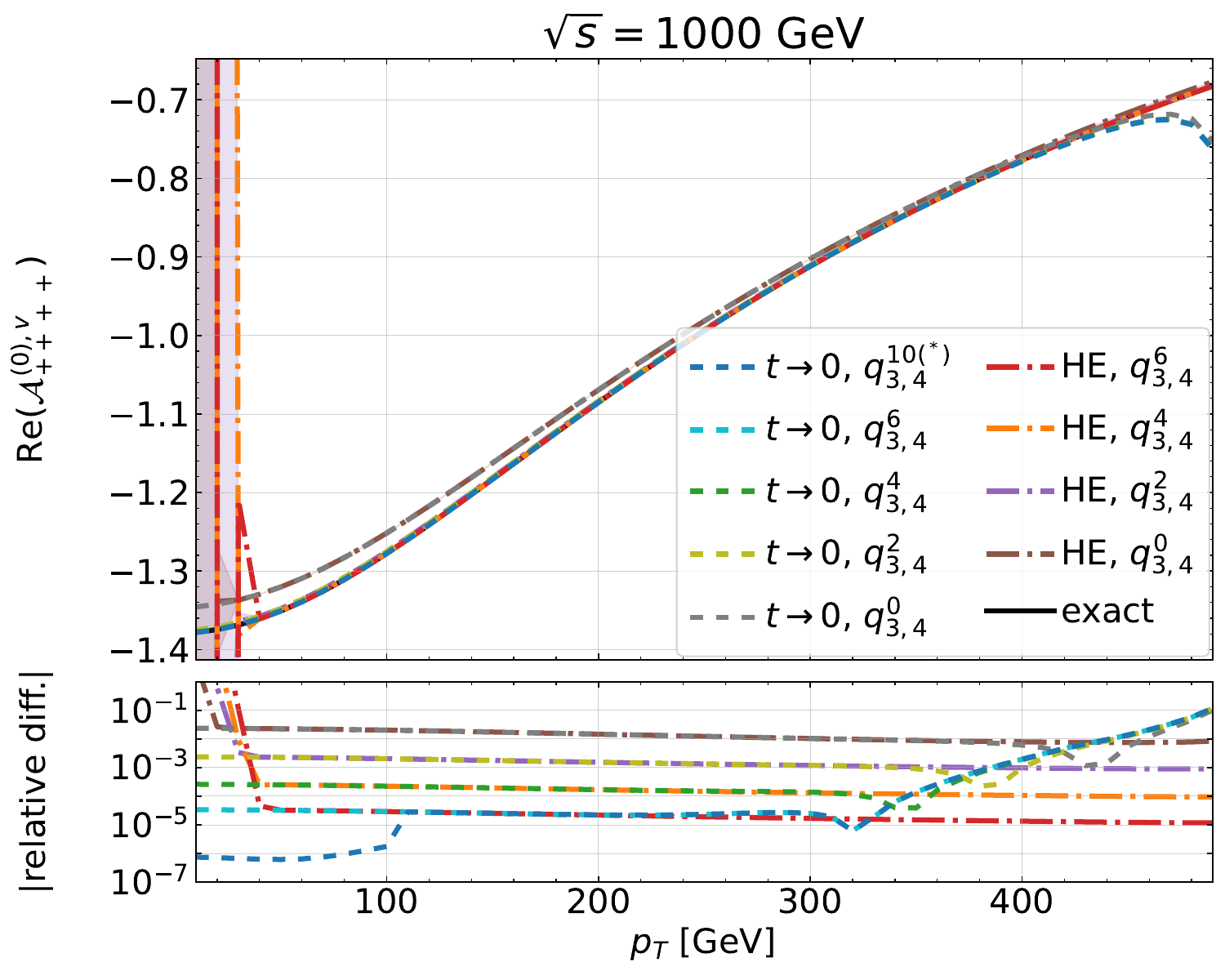} &
    \includegraphics[width=.45\textwidth]{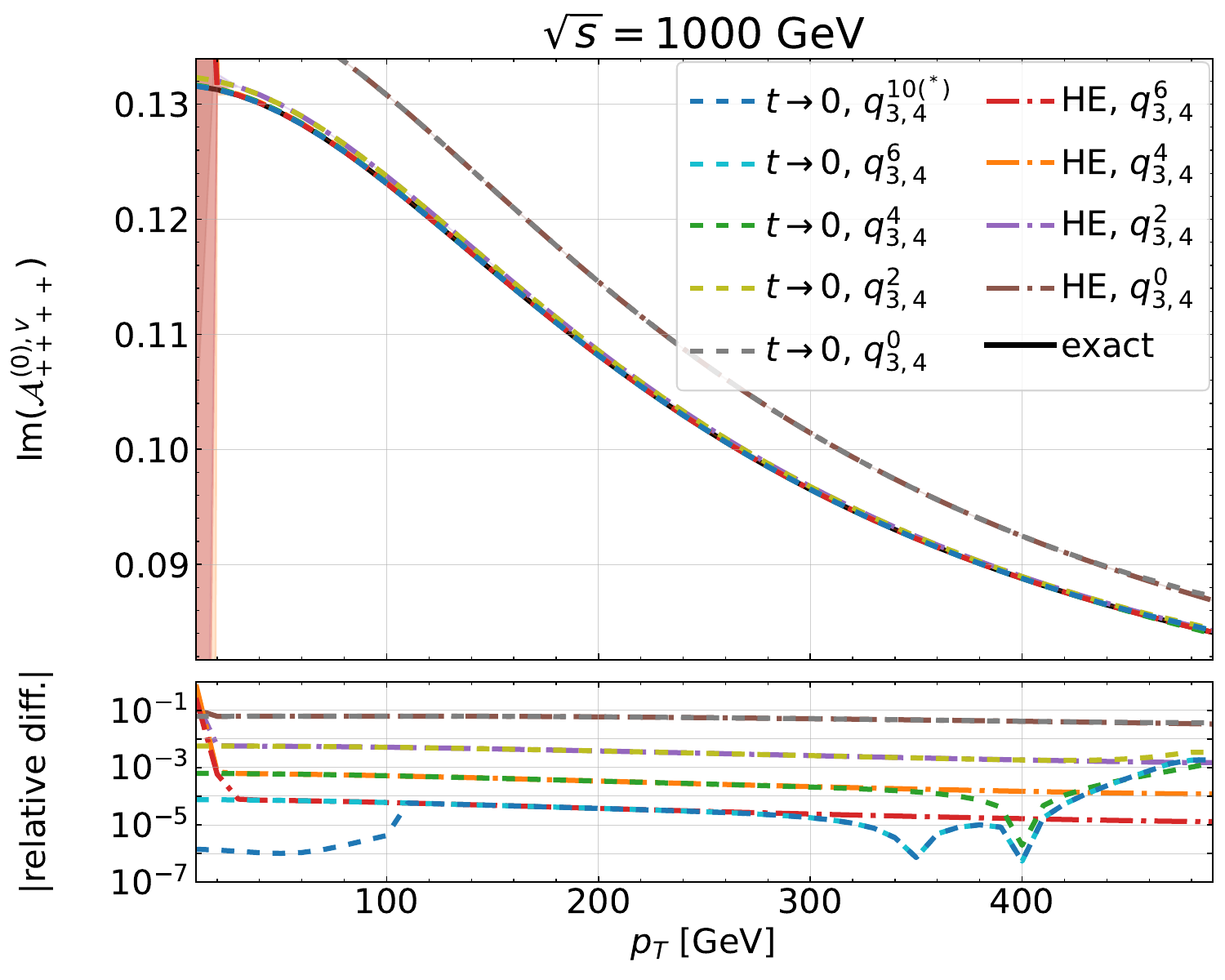} \\
    \includegraphics[width=.45\textwidth]{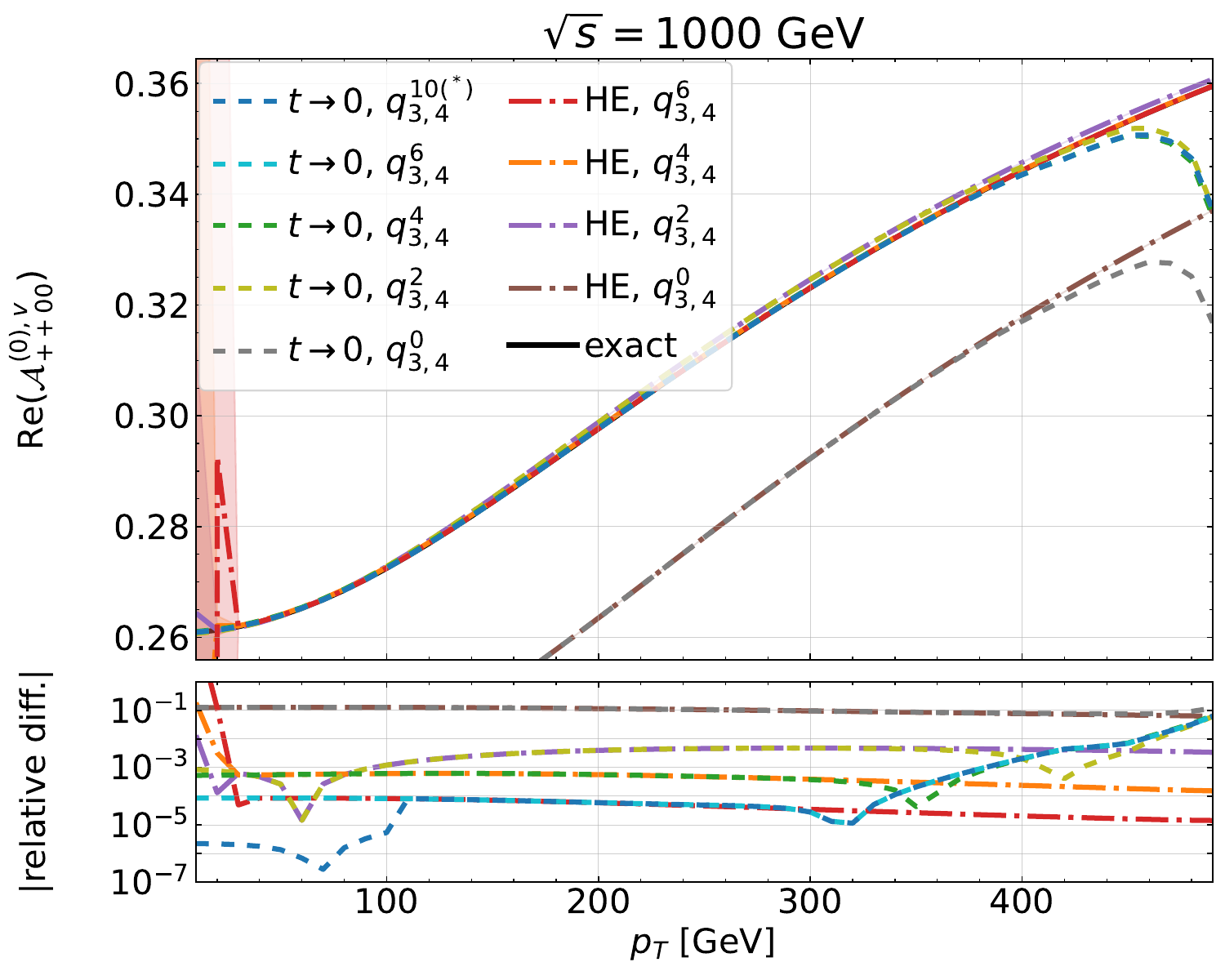} &
    \includegraphics[width=.45\textwidth]{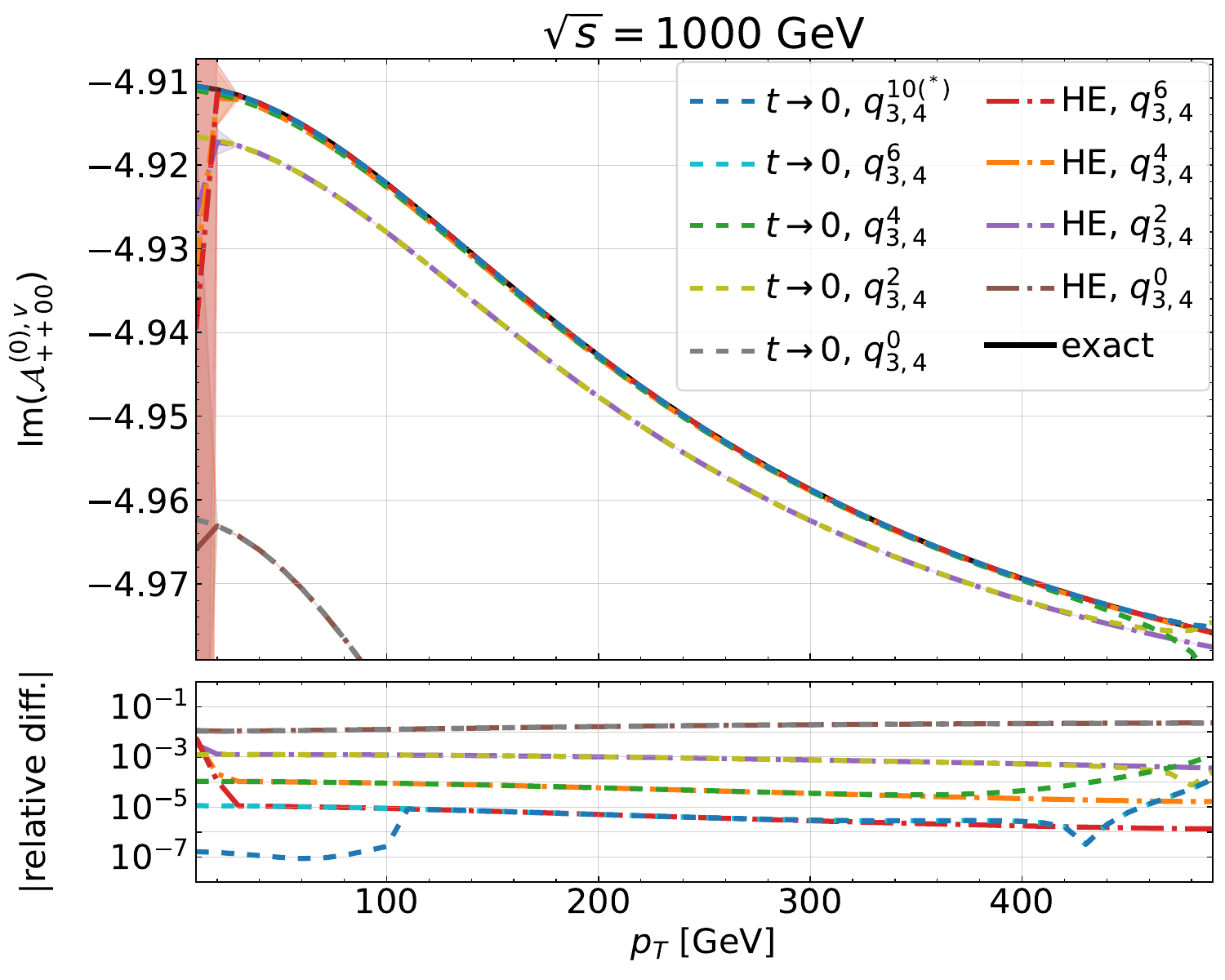}
    \end{tabular}
    \end{center}
  \caption{\label{fig::ampl1lv_sqrts1000}
  Real and imaginary parts of ${\cal A}^{(0),v}_{++++}$ and ${\cal A}^{(0),v}_{++00}$ as a function of $p_T$ for $\sqrt{s}=1000$~GeV and $q_3^2 = q_4^2 = m_W^2$. High-energy and $t\to 0$ expansions are shown, including mass corrections up to $q_{3,4}^{\{0,2,4,6\}}$. Also shown are higher mass corrections up to $q_{3,4}^{10}$ for the $t\to 0$ expansion with fewer expansion terms in $t$ according to $t^{n_t}(q_3^2)^{n_3}(q_4^2)^{n_4}$ with $n_t+n_3+n_4\leq 5$ denoted by $\star$. Lower panels display the relative difference with respect to the exact result.}
\end{figure}

As an example, we discuss the real and imaginary parts of the two
helicity amplitudes ${\cal A}^{(0),v}_{++++}$ and ${\cal A}^{(0),v}_{++00}$, which are shown
in Fig.~\ref{fig::ampl1lv_sqrts1000} as a function of $p_T$ for $\sqrt{s}=1000$~GeV. Each panel displays several
coloured curves, with the dashed and dash-dotted curves corresponding to the
$t$-expansion and the high-energy expansion, respectively; the exact
result is shown as a solid black line.\footnote{The black lines are barely visible since they are essentially covered by the coloured lines.} The lower panels of each figure
show the relative deviation from the exact results, using the same
line types and colours as in the upper panels. For each expansion-based
curve, we include the maximum number of expansion terms discussed
above but vary the expansion depth in $q_3^2$ and $q_4^2$ as indicated
in the legend. We also apply the aforementioned Padé improvement to
both the $t$-expansion and the high-energy expansion. 

Note that the expressions of the forward expansions are obtained from two different methods such that the expressions up to $q_{3,4}^6$ contain terms up to $t^{10}$, while the higher mass terms contain only a limited number of terms in the $t$-expansion following $t^n (q_3^2)^m (q_4^2)^k$ with $4,5\geq n+m+k>3$ depending on the helicity amplitude. The limited number of terms in $t$ lead to the side-effect, that the inclusion of those terms reduces significantly the range of convergence of the constructed Pad\'e approximations. Therefore, we include these higher mass terms dynamically only if $p_T\leq100$~GeV.

Let us begin the discussion with the $t$-expansion which is valid for lower $p_T$ values. Indeed, we observe that the $t$-expansion agree with the exact result up to approximately $p_T =
300$~GeV with a precision below $0.01\%$. Below $p_T = 100$~GeV the accuracy is improved by about two orders of magnitude due to the additional terms of the mass expansion. Deviations appear only
at the right edge of the plots.

The high-energy expansion exhibits analogous behaviour: For large
values of $p_T$, there is perfect agreement with the exact result,
achieving a precision of $4$ to $5$ significant digits when the sixth-order
terms in the expansion in $q_3^2$ and $q_4^2$ are taken into
account. Only at low values of $p_T$ does the Pad\'e-improved
high-energy expansion begin to diverge, a fact reflected in the large
uncertainty estimate associated with our Pad\'e method. This is evident
from the red band covering the entire range of the $y$-axis.

Regarding the expansion in terms of external masses, it is instructive
to focus on the lower panels. One observes that each new expansion term yields a gain of a factor of approximately
10, leading to an agreement with the exact result to within
four or five significant figures. Furthermore, in all four panels
shown in Fig.~\ref{fig::ampl1lv_sqrts1000}, it can be seen that the
$t$-expansion and the high-energy expansion -- incorporating the same
terms for the external masses -- agree over a wide range of intermediate
$p_T$ values.  As previously mentioned, the $t$-expansion, which
includes more terms in $q_3^2$ and $q_4^2$ but fewer terms in
$t^n$, yields better results below $p_T \approx 100$~GeV.

\begin{figure}[t]
    \begin{center}
    \begin{tabular}{cc}
    \includegraphics[width=.45\textwidth]{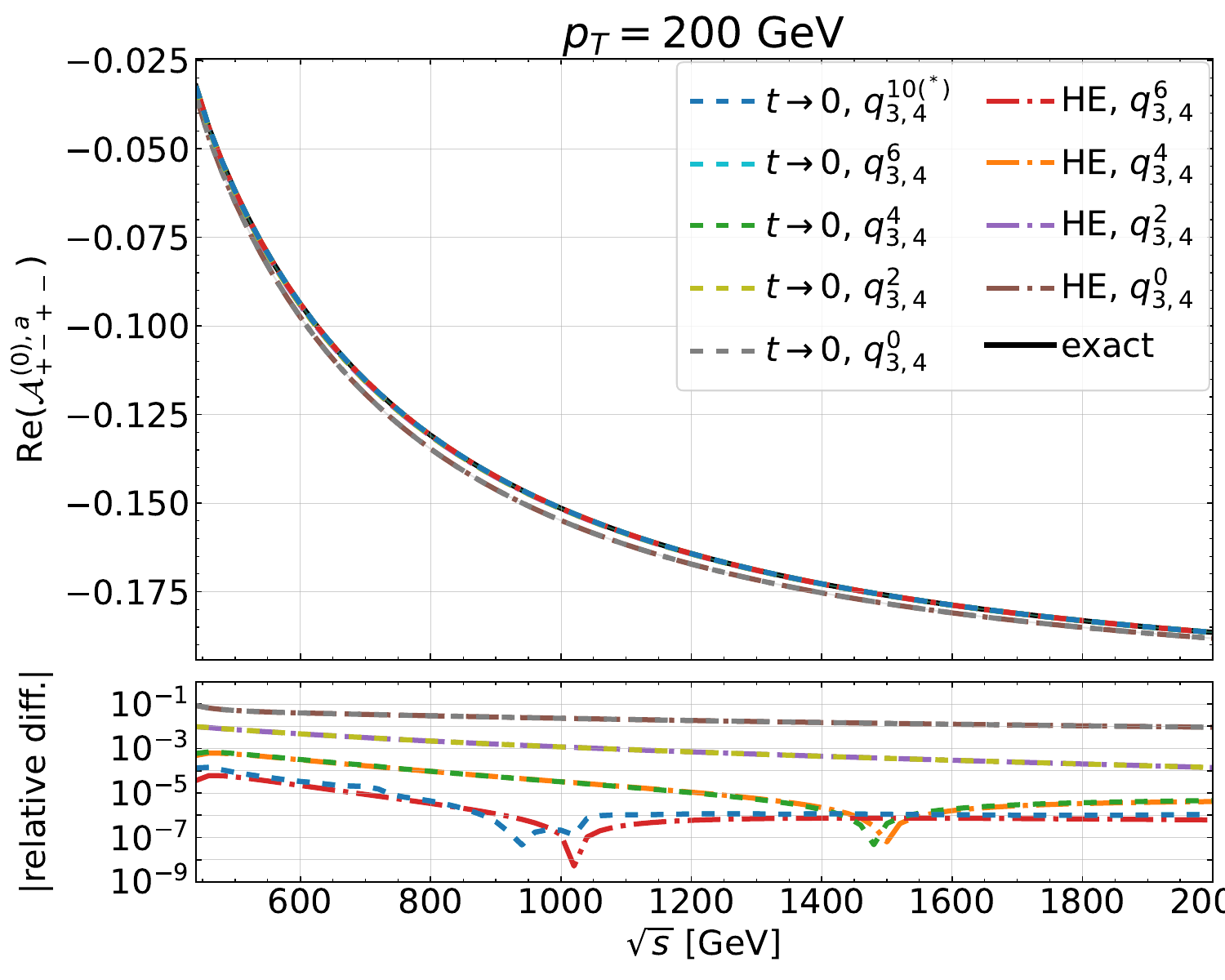} &
    \includegraphics[width=.45\textwidth]{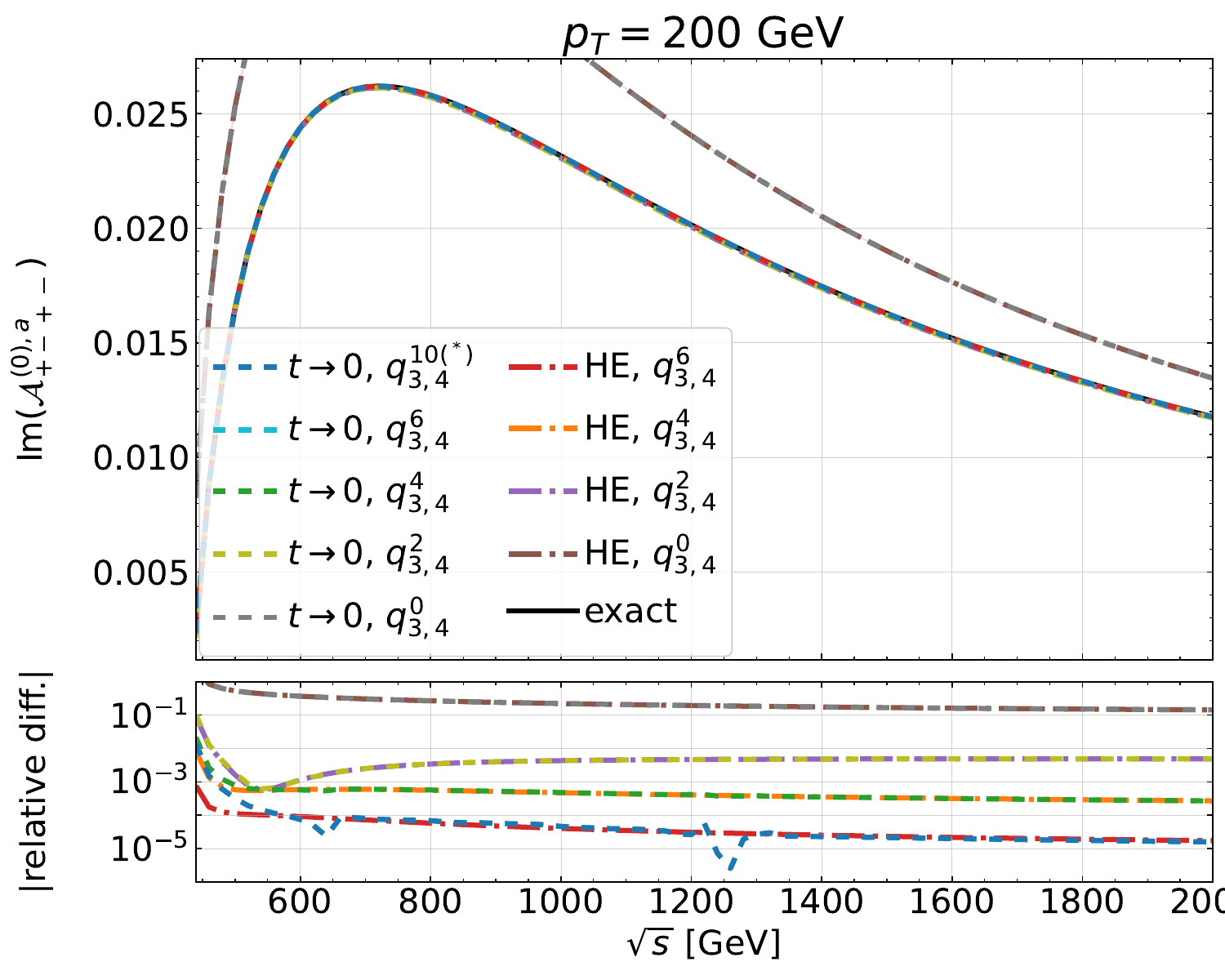} \\
    \includegraphics[width=.45\textwidth]{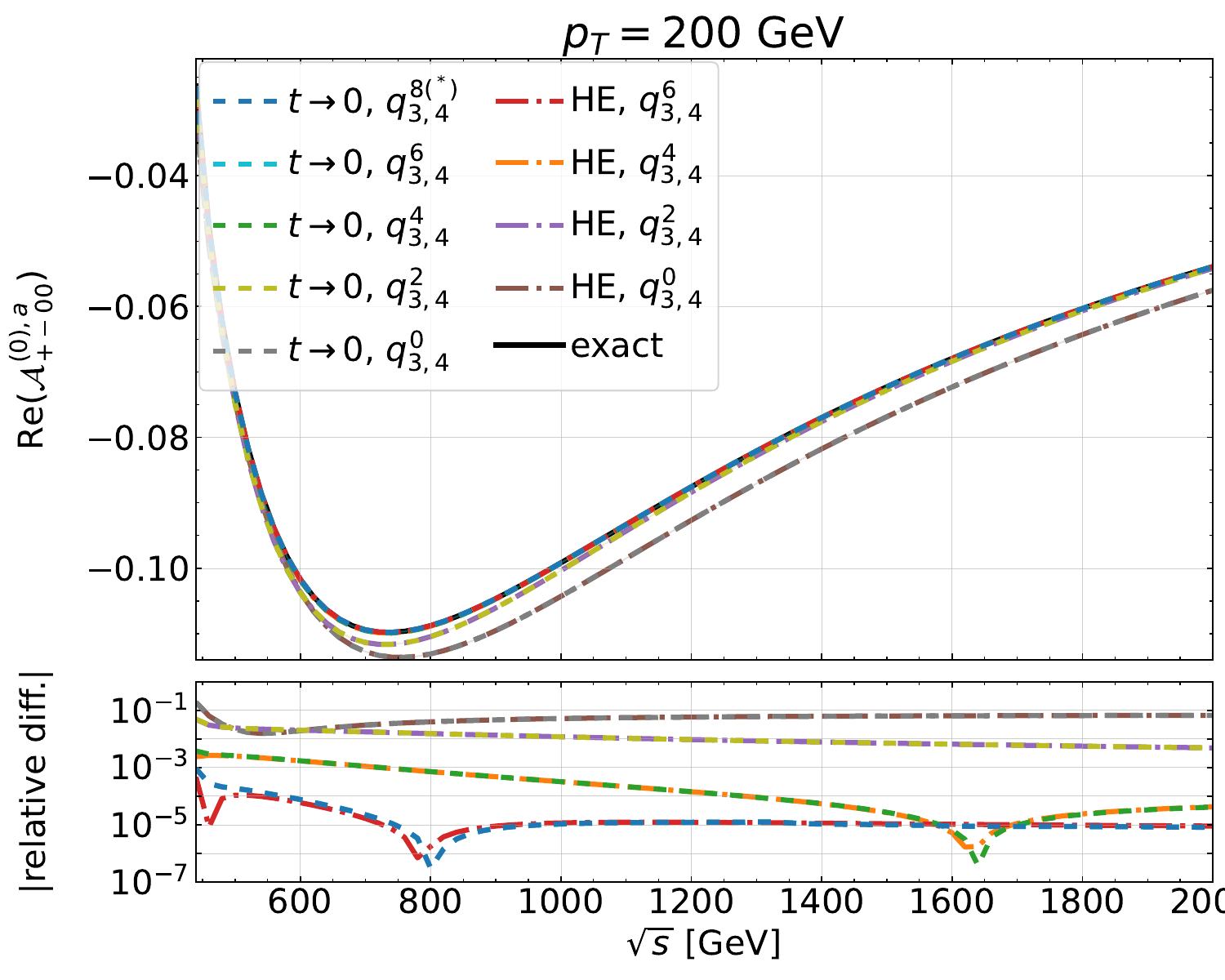} &
    \includegraphics[width=.45\textwidth]{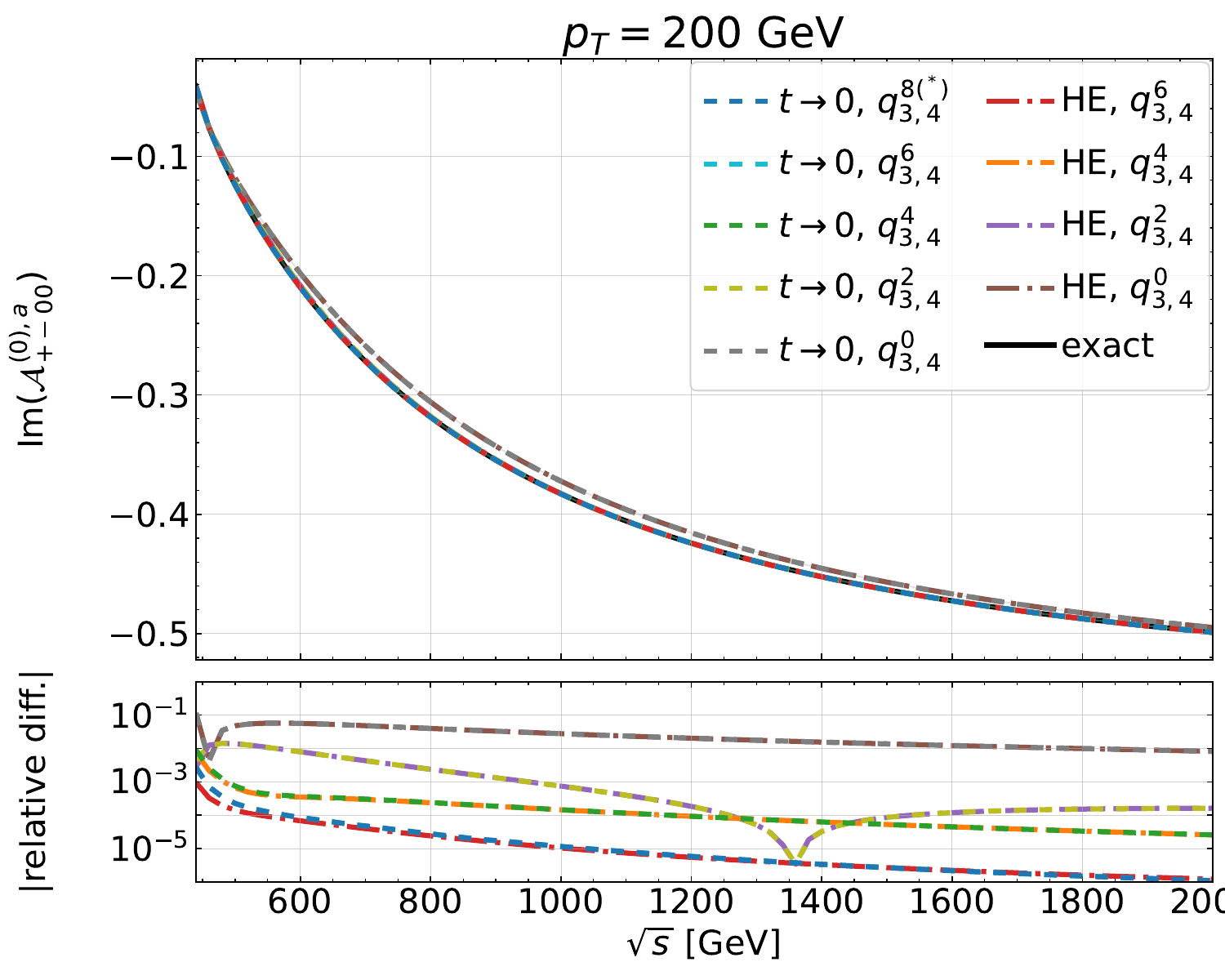}
    \end{tabular}
    \end{center}
  \caption{\label{fig::ampl1la_pt200}
  Same as Figure \ref{fig::ampl1lv_sqrts1000} but for ${\cal A}^{(0),a}_{+-+-}$ and ${\cal A}^{(0),a}_{+-00}$ and $p_T=200$~GeV as a function of $\sqrt{s}$.}
\end{figure}

Figure~\ref{fig::ampl1la_pt200} shows the helicity amplitudes ${\cal A}^{(0),a}_{+-+-}$ and ${\cal A}^{(0),a}_{+-00}$, plotted against $\sqrt{s}$ at a
fixed $p_T=200$~GeV.  We use the same colours and line styles for the
various expansions and the exact results.  The quality of the
expansions appears to be essentially the same as in
Figure~\ref{fig::ampl1lv_sqrts1000}: Including additional terms in the
expansion in external masses improves the agreement with the exact
result, which can be reproduced with an accuracy of at least three
significant digits.  Furthermore, we observe good agreement between
the $t$-expansion and the high-energy expansion, suggesting that we
can cover the entire phase space by combining both expansions.
Some curves in Fig.~\ref{fig::ampl1la_pt200} show a peak-like structure in the lower panels. This is related to crossing of the respective expansion and the exact result leading to small numbers in the numerator of the relative difference.

\subsection{\label{sub::2loop}Two-loop results}

\begin{figure}[t]
    \begin{center}
    \begin{tabular}{cc}
    \includegraphics[width=.45\textwidth]{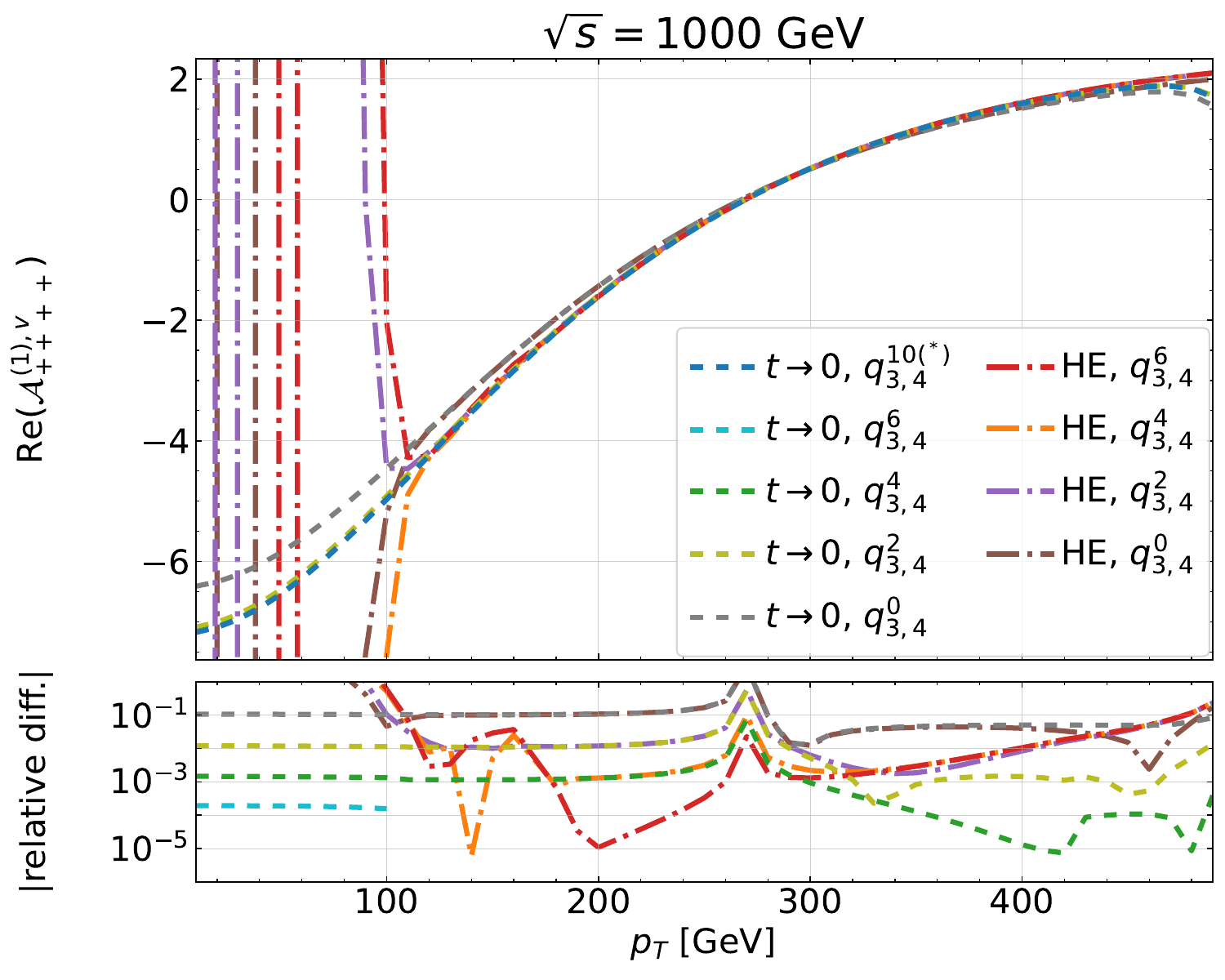} &
    \includegraphics[width=.45\textwidth]{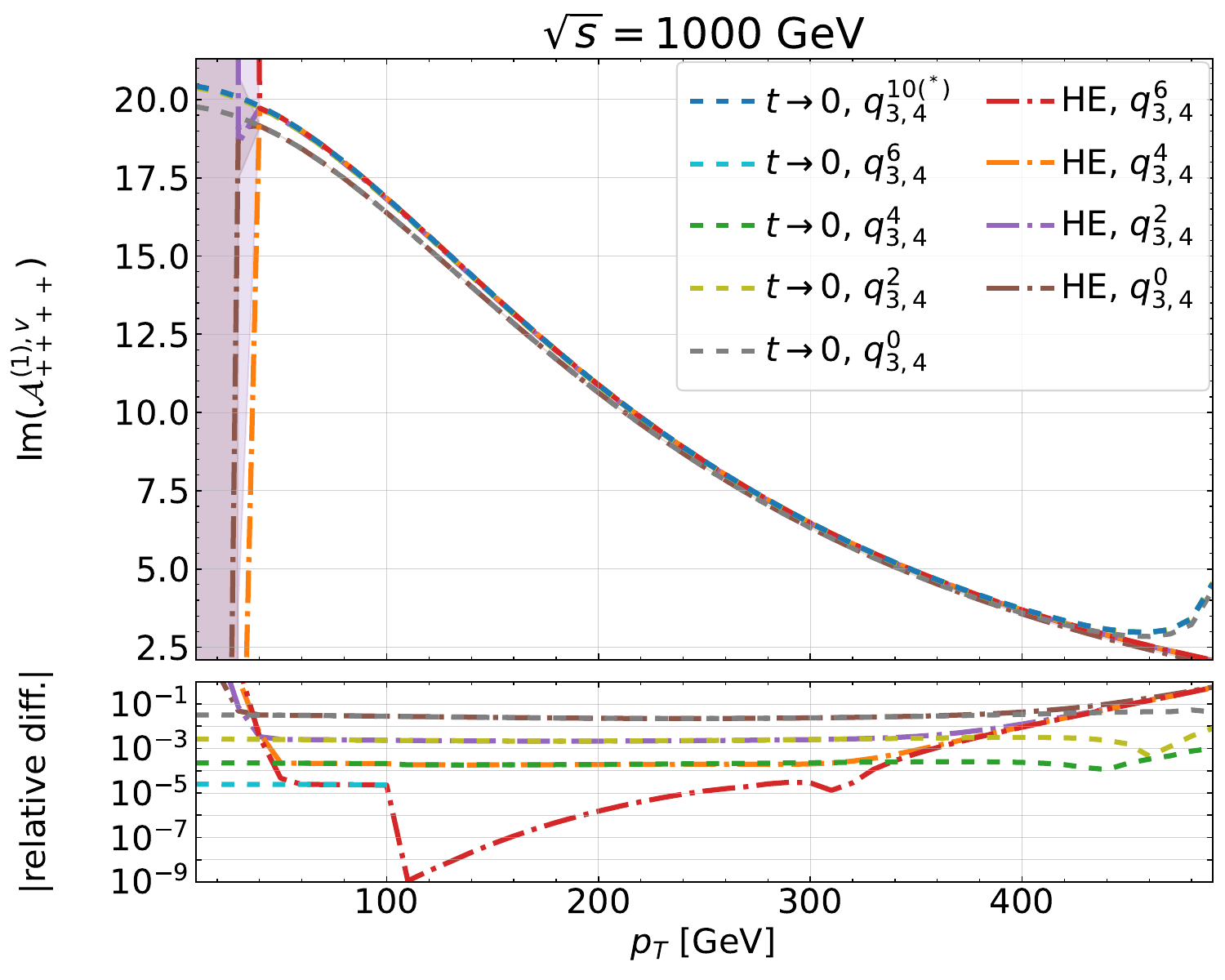} \\
    \includegraphics[width=.45\textwidth]{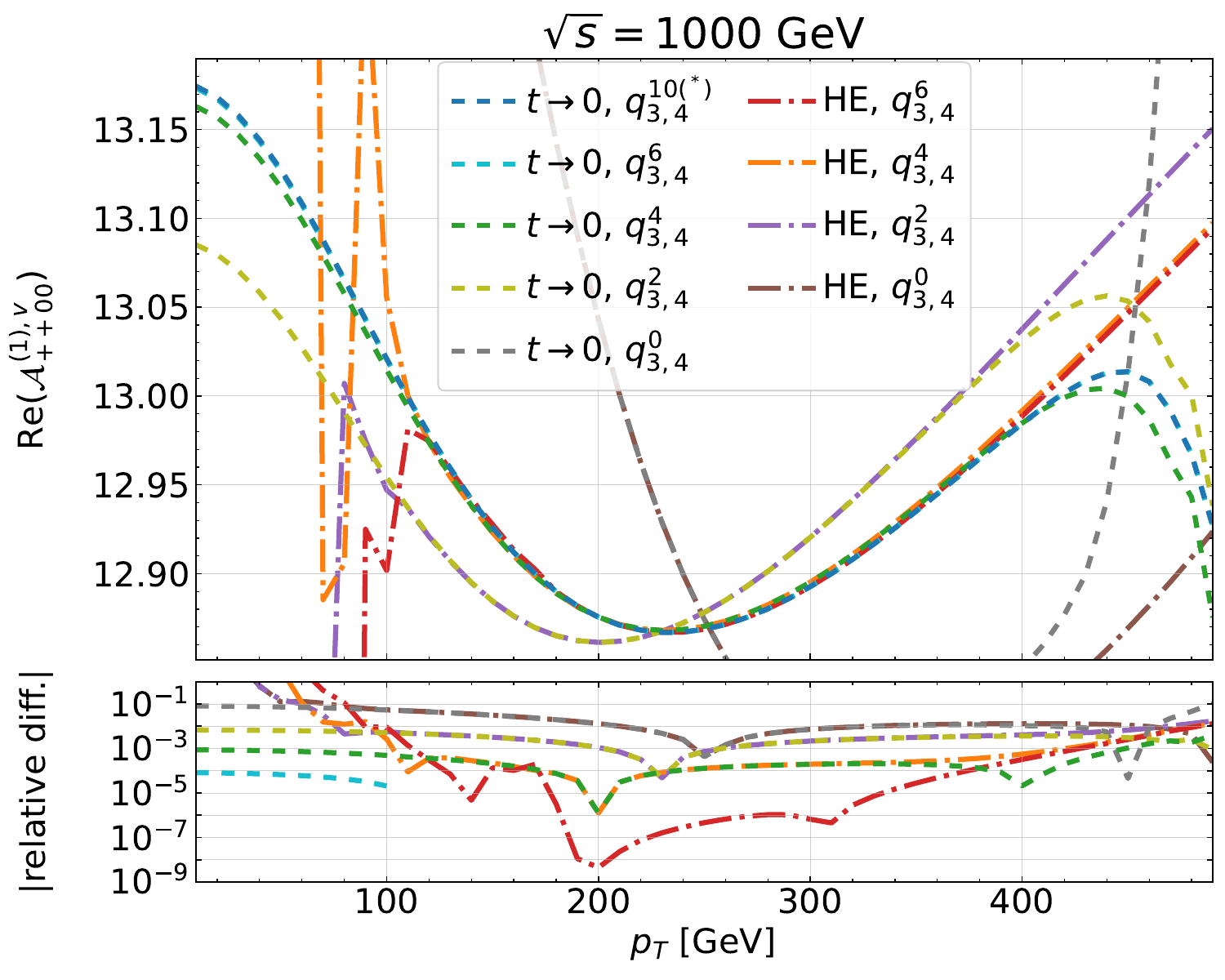} &
    \includegraphics[width=.45\textwidth]{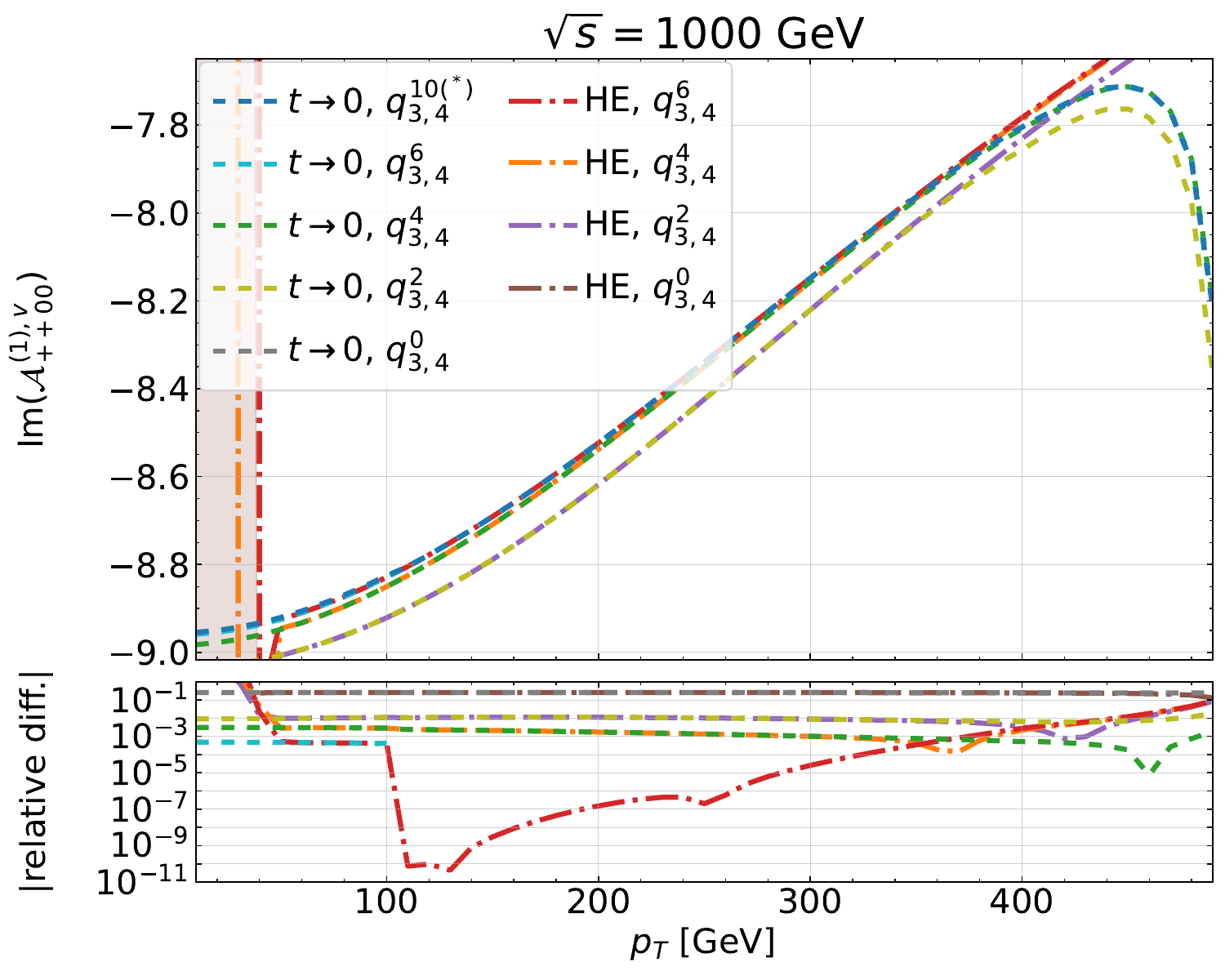}
    \end{tabular}
    \end{center}
  \caption{\label{fig::ampl2lv_sqrts1000}
  Real and imaginary parts of ${\cal A}^{(1)}_{++++}$ and ${\cal A}^{(1)}_{++00}$ as a function of $p_T$ for $\sqrt{s}=1000$~GeV and $q_3^2 = q_4^2 = m_W^2$. High-energy and $t\to 0$ expansions are shown, including mass corrections up to $q_{3,4}^{\{0,2,4,6\}}$. Also shown are higher mass corrections up to $q_{3,4}^{10}$ for the $t\to 0$ expansion with fewer expansion terms in $t$ according to $t^{n_t}(q_3^2)^{n_3}(q_4^2)^{n_4}$ with $n_t+n_3+n_4\leq 5$ denoted by $\star$. Lower panels display the relative difference with respect to the best approximation of the $t\to 0$ expansion.}
\end{figure}

\begin{figure}[t]
    \begin{center}
    \begin{tabular}{cc}
    \includegraphics[width=.45\textwidth]{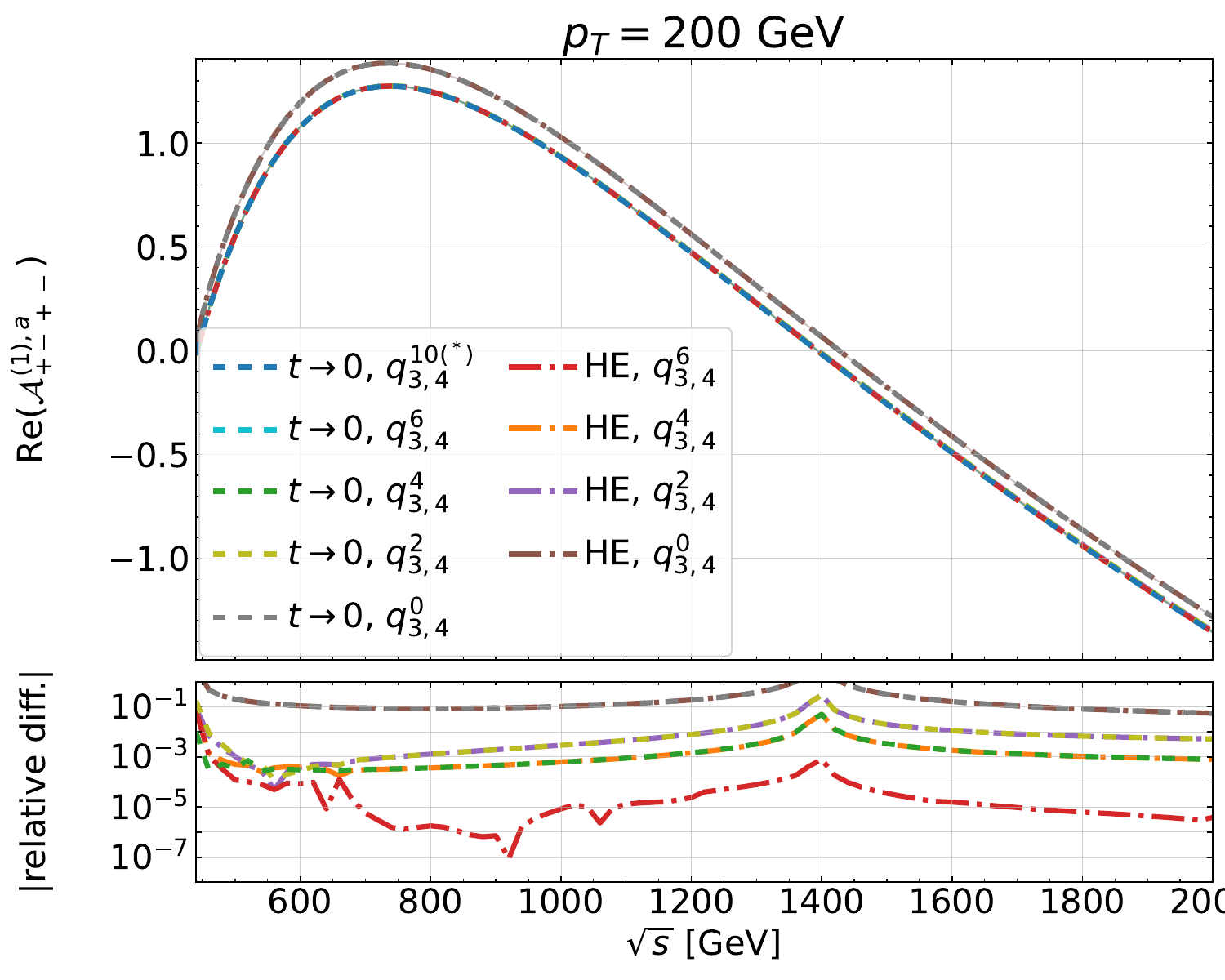} &
    \includegraphics[width=.45\textwidth]{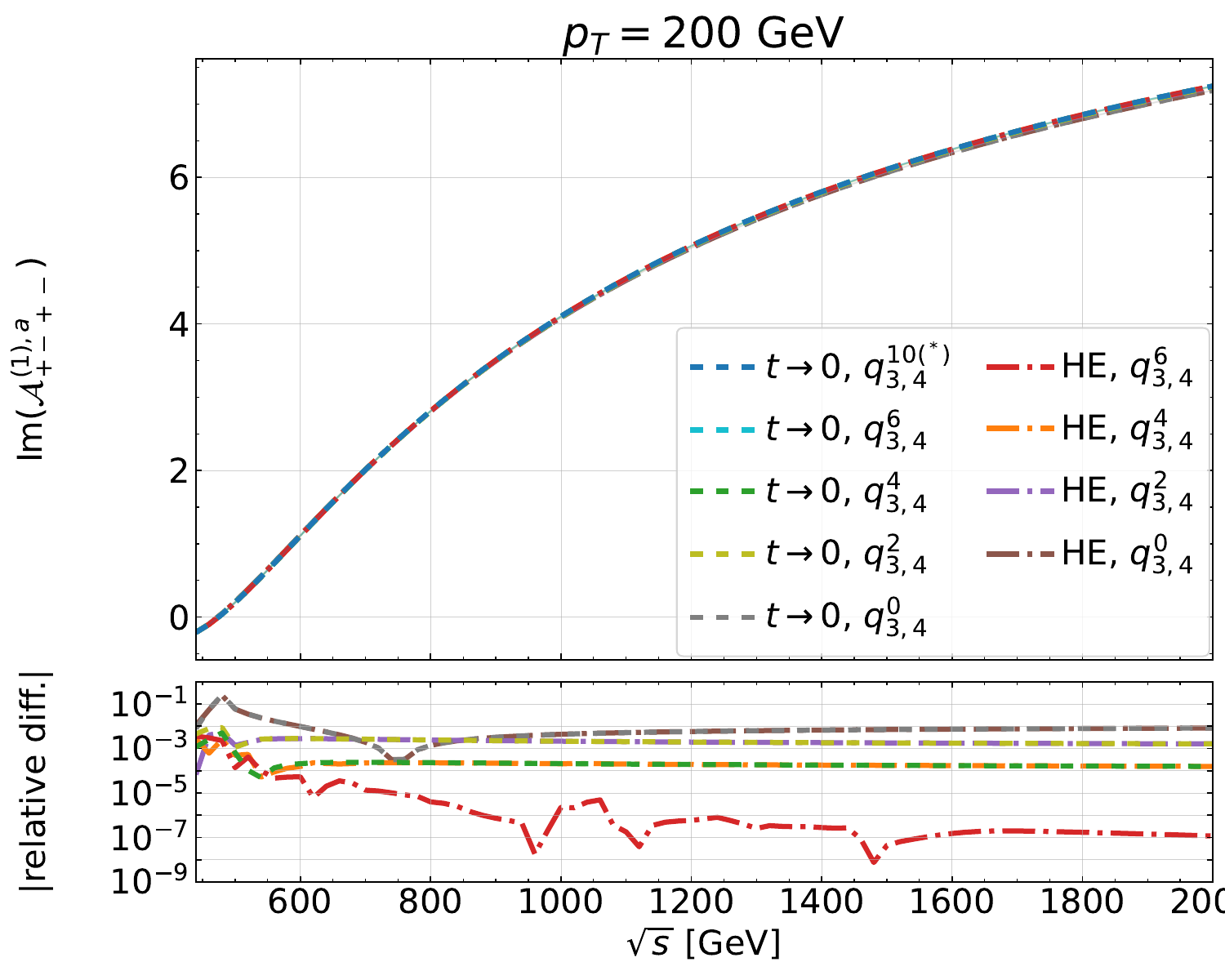} \\
    \includegraphics[width=.45\textwidth]{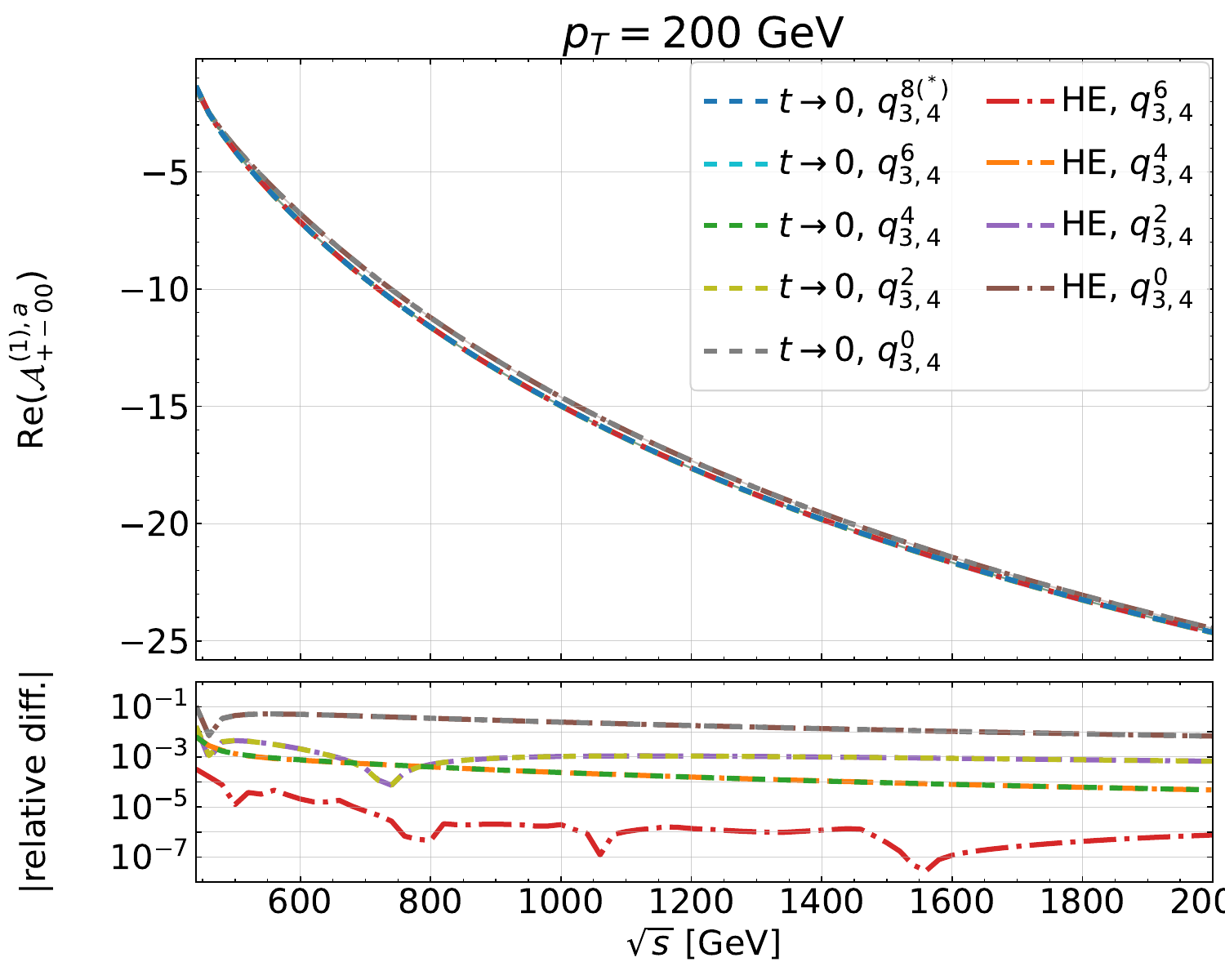} &
    \includegraphics[width=.45\textwidth]{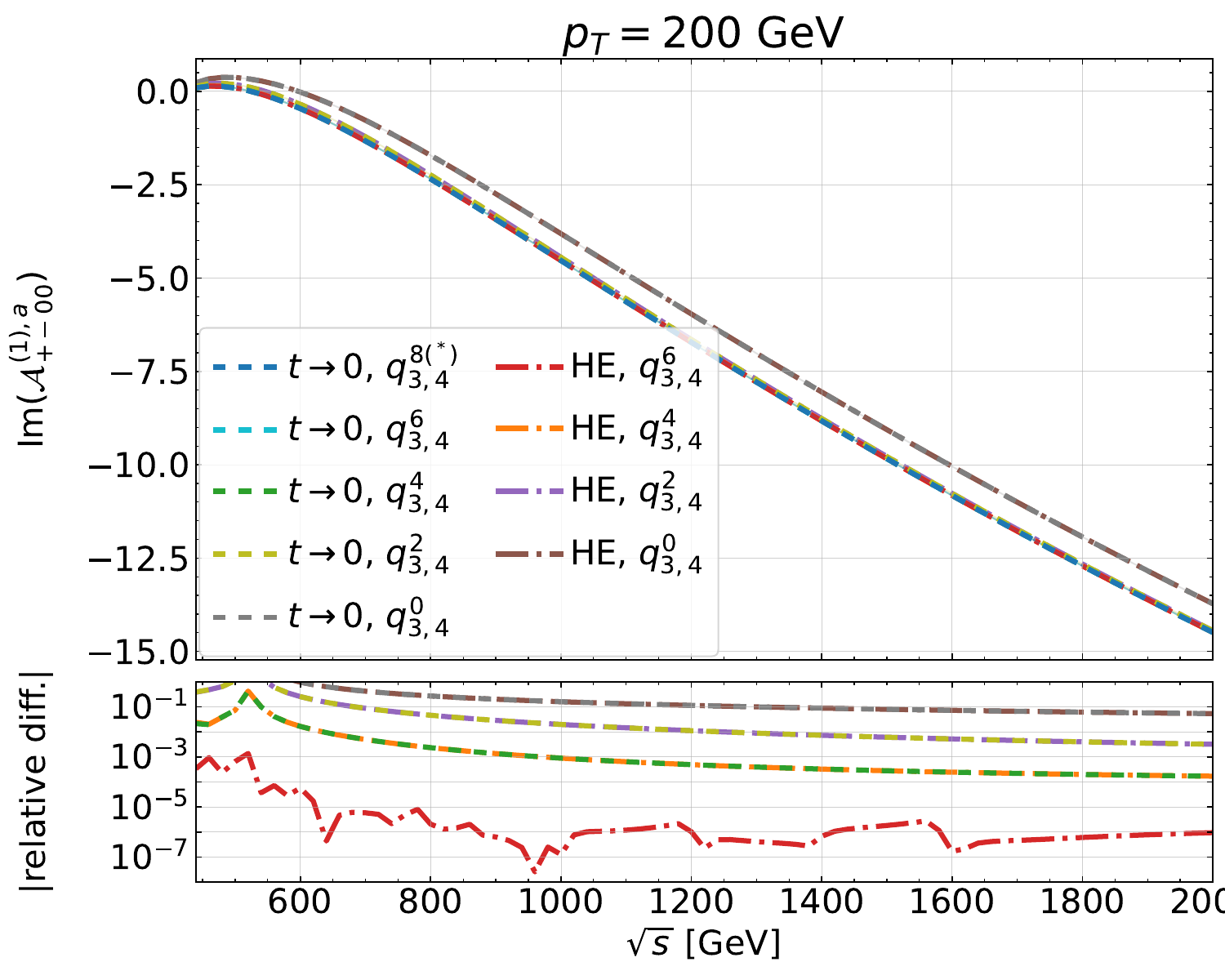}
    \end{tabular}
    \end{center}
  \caption{\label{fig::ampl2la_pt200}
  Same as Figure \ref{fig::ampl2lv_sqrts1000} but for ${\cal A}^{(1),a}_{+-+-}$ and ${\cal A}^{(1),a}_{+-00}$ and $p_T=200$~GeV as a function of $\sqrt{s}$.}
\end{figure}

In Figs.~\ref{fig::ampl2lv_sqrts1000} and \ref{fig::ampl2la_pt200},
we present the two-loop results for the same helicity amplitudes as in
the previous subsection, specifically for fixed values of
$\sqrt{s}=1000$~GeV and $p_T=200$~GeV, respectively. Since no exact
result is available, we normalize the results in the lower panels to
the best approximation derived from the $t$-expansion. Consequently,
the relative differences in Fig.~\ref{fig::ampl2lv_sqrts1000} tend
to increase at higher $p_T$ values. This behaviour arises because the
$t$-expansion loses accuracy in this region of phase space. We do not
observe such behaviour in Fig.~\ref{fig::ampl2la_pt200}, as both
expansions yield good results at $p_T=200$~GeV, even for higher values
of $\sqrt{s}$.

Fig.~\ref{fig::ampl2lv_sqrts1000} clearly shows that the high-energy expansion is no longer valid for smaller values of $p_T$. For some
helicity amplitudes, significant deviations from the $t$-expansion are
visible at $p_T \approx 100$~GeV; in other cases, the deviation only
appears at lower $p_T$ values. No such effect is
observed in Fig.~\ref{fig::ampl2la_pt200}, as both expansions yield
good results for all $\sqrt{s}$ values at the chosen $p_T$. However,
it is evident that the relative differences worsen slightly as
$\sqrt{s}$ decreases where the high-energy expansion is less precise.

Next, let us consider the regions of the plots that are crucial to our
study -- namely, the combination of the two expansions at $p_T\sim 200$~GeV. In all cases, we
find perfect agreement between the $t$-expansion and the high-energy
expansion; in most instances, the precision even exceeds five
significant figures. Furthermore, we observe a visible shift
after including the quadratic terms in $q_3^2$ and
$q_4^2$. By contrast, the additional fourth- and sixth order terms result in only slight shifts, which is barely noticeable in the upper panels. Similarly to the one-loop amplitudes, each additional order in the mass expansion improves the accuracy by about a factor of 10, so that the differences between the $q_{3,4}^4$ and $q_{3,4}^6$ terms are about $0.1\%$ indicating a similar precision for our best prediction.
In Section~\ref{sec::masters}, we find agreement with the exact
numerical result at the 1\% level. Our results in this section indicate
that expansions at the amplitude level lead to better results.

Some of the curves in the lower panels exhibit a cusp-like
structure. This is related to the intersections of the
respective approximation and the best approximation of the
$t\to0$ expansion, which serves as a normalization.

We have presented two types of plots in which we fix either $\sqrt{s}$
or $p_T$ and vary the other parameter. It should be noted that for
lower values of fixed $\sqrt{s}$, the kinematically allowed
maximum value of $p_T$ is lower, causing the $t$-expansion to work
even better. On the other hand, at higher values of $\sqrt{s}$, we
reach deeper into the validity range of the high-energy
expansion. With $p_T$ fixed at 200~GeV, both the $t$-expansion and the
high-energy expansion cover the entire range of $\sqrt{s}$ values. For
smaller $p_T$ values, the $t$-expansion performs even better, whereas
for larger $p_T$ values, the convergence of the high-energy expansion
improves. Therefore, we are confident that the combination of the
forward and high-energy expansions covers the entire physical phase
space for $W^+W^-$ pair production.

\begin{figure}[t]
    \begin{center}
    \begin{tabular}{cc}
    \includegraphics[width=.45\textwidth]{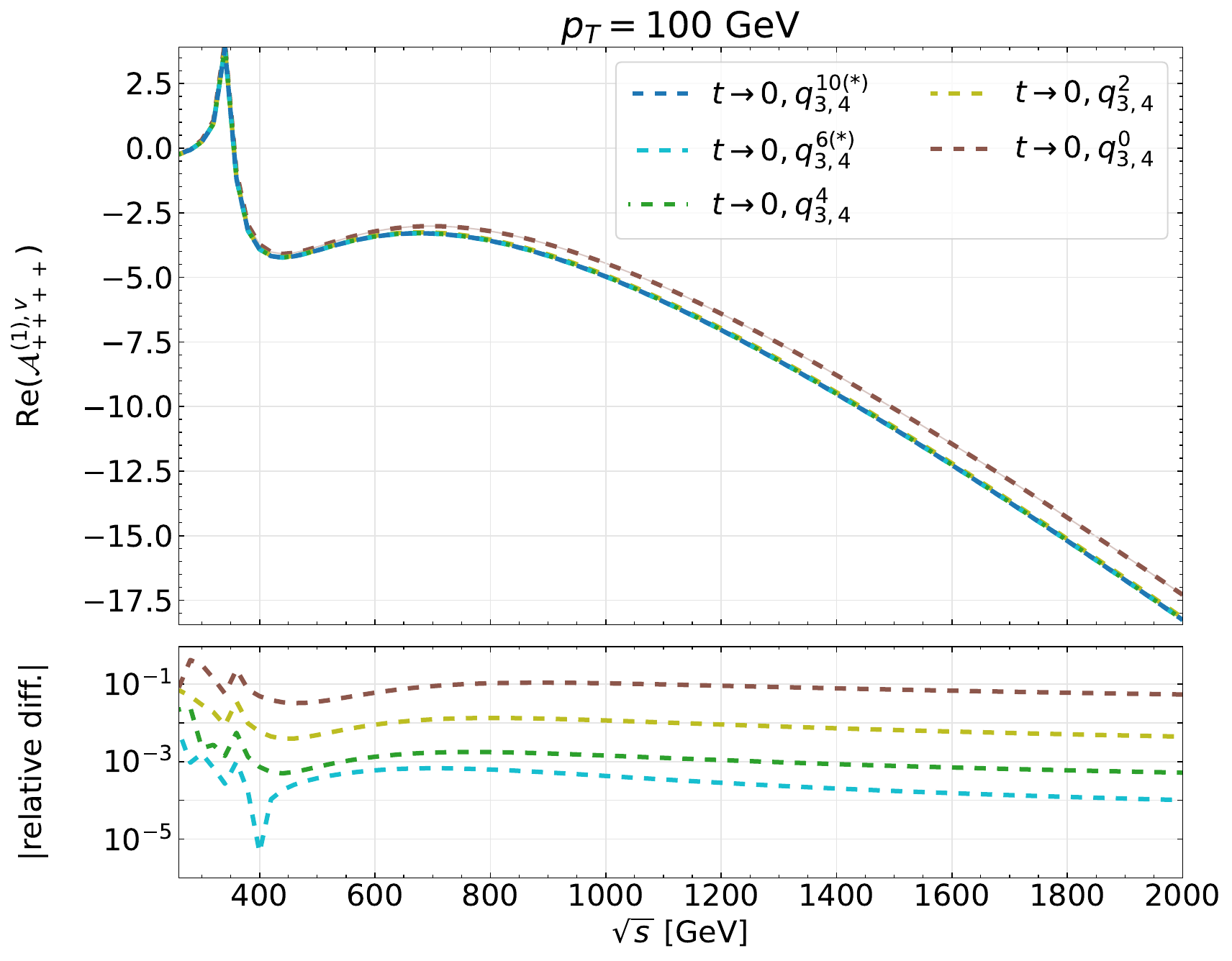} &
    \includegraphics[width=.45\textwidth]{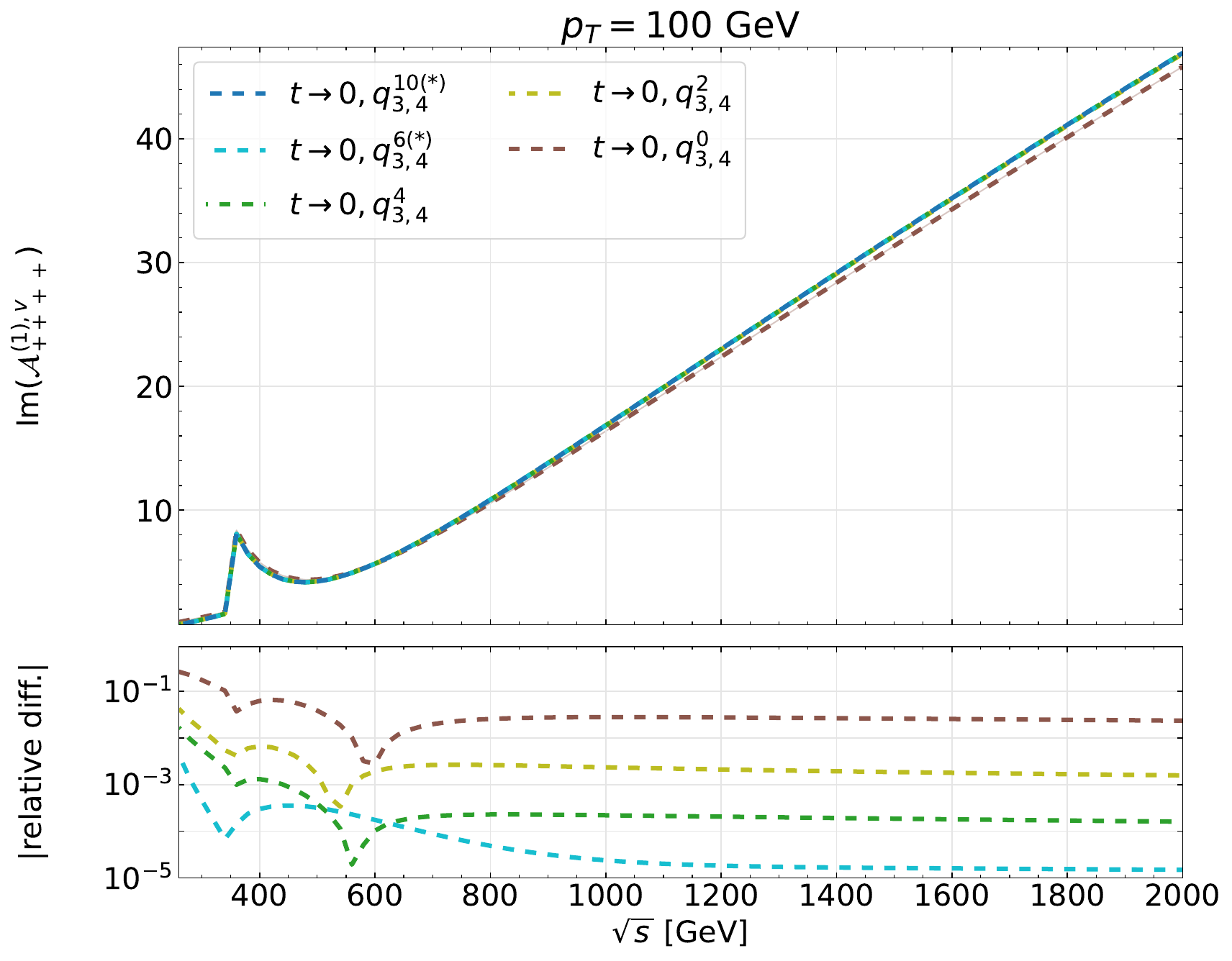} 
    \end{tabular}
    \end{center}
  \caption{\label{fig::hel1111}
  Real and imaginary parts of ${\cal A}^{(1),v}_{++++}$ for $p_T=100$~GeV as a function of $\sqrt{s}$.}
\end{figure}

Finally, in Fig.~\ref{fig::hel1111}, we consider an example of a
helicity amplitude that demonstrates the power of our approach. We
choose $p_T=100$~GeV and vary $\sqrt{s}$. For this region of phase
space, the forward expansion suffices to achieve sub-per-mille
accuracy, as can be seen in the lower panels of
Fig.~\ref{fig::hel1111}. In the upper panels, a pronounced cusp is
visible at $\sqrt{s}=2 m_t$. It arises because the virtual top-quark
pair can be produced as an $S$-wave state in this spin configuration,
leading to an enhancement of the cross section. The
\textit{expand and match} method we use to calculate the forward expansion
incorporates a generalized expansion around the threshold $\sqrt{s}=2
m_t$. This includes non-analytic terms responsible for the kink
visible in both the real and imaginary parts of ${\cal
A}^{(1),v}_{++++}$. Thus, such structures are already captured in our
semi-analytic results, eliminating the need to resolve them through
numerous numerical evaluations near the $\sqrt{s}$ values where the
helicity amplitudes exhibit strong fluctuations.

\section{\label{sec::concl}Conclusions}

In this work, we present semi-analytical results for another $2\to 2$
process: the top-quark-induced contribution to $W$-boson pair
production via gluon fusion. To date, this process has been
investigated only using a numerical
approach~\cite{Bronnum-Hansen:2020mzk} and a      
grid of points which could be used by other research groups for phenomenological studies was not provided.

Our approach is based on expansions around the forward limit, which is
the most important from a phenomenological standpoint, the high-energy
limit (which also covers a significant portion of phase space), and an
expansion for large top-quark mass. 

Similar processes have previously been investigated in this
context. However, in all those cases, the final-state bosons were
electrically neutral and only the top quark appeared in the loop
integral. In $W^+W^-$ production, by contrast, bottom quarks are also
present in the loop diagrams, and the massive particle line is not
continuous. This leads to different master integrals. Since there are
fewer massive lines, it is indeed possible to obtain analytical
results in both the high-energy and forward limits without
encountering elliptic sectors. 

For the numerical evaluation, we supplement our expressions for the
helicity amplitudes with Pad\'e approximations, thereby covering the
entire phase space. Although our analytical expressions have a total
size of $3$~GB in zipped format, they possess a simple analytical structure and are
thus easily evaluated numerically. The analytical results for the
master integrals and the helicity amplitudes are available from the
authors upon request.

In the next step, we will implement our results into the \texttt{C++}
library \texttt{ggxy}~\cite{Davies:2025qjr}.  This will be done in conjunction with
the helicity amplitudes for $gg\to ZZ$ calculated in Ref.~\cite{Davies:2025out},
as well as the massless contributions~\cite{Gehrmann:2015ora,vonManteuffel:2015msa}.  In this way, we
obtain fast and flexible code for calculating the corresponding
gluon-induced cross sections, both at the partonic and hadronic
levels. Furthermore, it will be possible to access the individual helicity amplitudes, which can then also be used by other groups for further purposes.

\section*{Acknowledgements}

This work was supported by the Deutsche Forschungsgemeinschaft (DFG, German
Research Foundation) under grant 396021762 --- TRR 257 ``Particle Physics
Phenomenology after the Higgs Discovery''.
We thank Dominik Grau, Joshua Davies, Kay Sch\"onwald, Marco Vitti and Hantian Zhang for helpful discussions and useful comments on the manuscript, Micha\l{} Czakon for sharing with us an unpublished code for resumming residue series of Mellin-Barnes integrals,
Dominik Grau for generating optimized \texttt{FORM} files for the integral families and Kay Sch\"onwald for providing relations to perform the partial fraction decomposition for the forward expansion.

\begin{appendix}
\section{\label{app::sym}Symmetry relations among helicity amplitudes}

The symmetry relations among the helicity amplitudes for the
vector case are discussed in Refs.~\cite{Agarwal:2020dye,Davies:2025out}.
For the axial-vector case one obtains different relations
to reduce the 18 tensor structures of Eq.~(\ref{eq::Ti})
to the eight helicity amplitudes of Eq.~(\ref{eq::helamp}).
All helicity amplitudes with $\lambda_1=-$ can be obtained with the relation
\begin{equation}
\mathcal{A}^{a}_{\lambda_1\lambda_2\lambda_3\lambda_4}=-(-1)^{\delta_{\lambda_30}+\delta_{\lambda_40}}\mathcal{A}^{a}_{-\lambda_1 -\lambda_2 -\lambda_3 -\lambda_4}.
\end{equation}
Furthermore, we have the following relations for $q_3^2\leftrightarrow q_4^2$
\begin{equation}
\begin{array}{cclcccl}
\mathcal{A}^{a}_{++0-} &=& -\mathcal{A}^{a}_{++-0}\Big|_{q_3^2 \leftrightarrow q_4^2},
&\quad&
\mathcal{A}^{a}_{+-0-} &=& -\mathcal{A}^{a}_{+-+0}\Big|_{q_3^2 \leftrightarrow q_4^2},
\end{array}
\end{equation}
and for $\beta\to -\beta$
\begin{equation}
\begin{array}{cclcccl}
\mathcal{A}^{a}_{++--} &=& \mathcal{A}^{a}_{++++}\Big|_{\beta\to -\beta},
&\quad&
\mathcal{A}^{a}_{++-+} &=& \mathcal{A}^{a}_{+++-}\Big|_{\beta\to -\beta},\\
\mathcal{A}^{a}_{+--+} &=& \mathcal{A}^{a}_{+-+-}\Big|_{\beta\to -\beta},
&\quad&
\mathcal{A}^{a}_{+---} &=& \mathcal{A}^{a}_{+-++}\Big|_{\beta\to -\beta},\\
\mathcal{A}^{a}_{+++0} &=& -\mathcal{A}^{a}_{++-0}\Big|_{\beta\to -\beta},
&\quad&
\mathcal{A}^{a}_{++0+} &=& -\mathcal{A}^{a}_{++0-}\Big|_{\beta\to -\beta},\\
\mathcal{A}^{a}_{+-+0} &=& -\mathcal{A}^{a}_{+--0}\Big|_{\beta\to -\beta},
&\quad&
\mathcal{A}^{a}_{+-0+} &=& -\mathcal{A}^{a}_{+-0-}\Big|_{\beta\to -\beta},
\end{array}
\end{equation}
where $\beta$ is defined as
\begin{align}
\beta=\sqrt{1-2\frac{q_3^2+q_4^2}{s}+\frac{(q_3^2-q_4^2)^2}{s^2}},
\end{align}
which are used to obtain the remaining helicity amplitudes. Finally, we have checked that the following relations are satisfied by our results
\begin{equation}
\begin{array}{cclcccl}
\mathcal{A}^{a}_{++++} &=& -\mathcal{A}^{a}_{++++}\Big|_{q_3^2 \leftrightarrow q_4^2},
&\quad&
\mathcal{A}^{a}_{+-+-} &=& \mathcal{A}^{a}_{+-+-}\Big|_{q_3^2 \leftrightarrow q_4^2},\\
\mathcal{A}^{a}_{++00} &=& -\mathcal{A}^{a}_{++00}\Big|_{q_3^2 \leftrightarrow q_4^2},
&\quad&
\mathcal{A}^{a}_{+-00} &=& \mathcal{A}^{a}_{+-00}\Big|_{q_3^2 \leftrightarrow q_4^2},\\
\mathcal{A}^{a}_{+++-} &=& -\mathcal{A}^{a}_{+++-}\Big|_{q_3^2 \leftrightarrow q_4^2,\,\beta\to -\beta},
&\quad&
\mathcal{A}^{a}_{+-++} &=&\mathcal{A}^{a}_{+-++}\Big|_{q_3^2 \leftrightarrow q_4^2,\,\beta\to -\beta}.
\end{array}
\end{equation}
From these relations it directly follows that $\mathcal{A}^{a}_{++++}$ and $\mathcal{A}^{a}_{++00}$ vanish for $q_3^2=q_4^2$.

\section{\label{app::largemt}Large-$m_t$ expansion of the helicity amplitudes}

In this appendix we discuss the results for the 
helicity amplitudes in the large-$m_t$ limit. 
The radius of convergence is smaller than that of $ZZ$ production, for
example, which is due to a branch cut at $s=0$ arising from massless
bottom quarks.  In the case of $gg\to ZZ$, by contrast, the lowest
branch cut lies at $s=(2m_t)^2$.  Consequently, the utility for
phenomenological applications is limited.  However, due to their
simple structure, the analytical results are very compact and may serve
as cross-checks for other calculations. For instance, we
successfully verified our results for the forward expansions.

We provide analytic results for all
helicity amplitudes, calculated up to orders $1/m_t^{10}$ and $1/m_t^{4}$ at the one- and two-loop levels, respectively, in the supplementary material to this paper~\cite{progdata}.

\section{\label{app::topos}Topology definitions}
In this appendix we show the topology definitions of the two scalar integrals defined in Section~\ref{sec::masters} using the notation of {\tt Kira}:

{\small
\begin{verbatim}
   integralfamilies:
     - name: "ggwwfull2l8"
       loop_momenta: [p3, p7]
       propagators:
         - ["mts - p7^2", 0]
         - ["mts - p3^2", 0]
         - ["-(p3 + q3)^2", 0]
         - ["-(p3 + q1 + q3)^2", 0]
         - ["-(p3 - p7 - q2)^2", 0]
         - ["-(p3 - p7)^2", 0]
         - ["-(p7 + q1 + q2 + q3)^2", 0]
         - ["-(p3 + q2)^2", 0]
         - ["-(p7 + q3)^2", 0]

   - name: "ggwwfull2l11"
       loop_momenta: [p2, p6]
       propagators:
         - ["mts - p6^2", 0]
         - ["mts - p2^2", 0]
         - ["mts - (p2 - q1)^2", 0]
         - ["-(p2 + q2 + q3)^2", 0]
         - ["-(p2 - p6)^2", 0]
         - ["-(p2 - p6 + q2)^2", 0]
         - ["-(p6 + q3)^2", 0]
         - ["-(p2 + q3)^2", 0]
         - ["-(p6 + q1)^2", 0]
\end{verbatim}
}

\end{appendix}

\bibliographystyle{jhep}

\bibliography{inspire.bib,extra.bib,extra_linz.bib}

\end{document}